\documentclass[10pt,epsf,preprint]{aastex}
\usepackage{epsf}
\input psfig 

\begin{document}
\title{Early-type galaxies in the SDSS. II. Correlations between observables}

\author{
Mariangela Bernardi\altaffilmark{\ref{Chicago},\ref{CMU}},
Ravi K. Sheth\altaffilmark{\ref{Fermilab},\ref{Pitt}},
James Annis\altaffilmark{\ref{Fermilab}},
Scott Burles\altaffilmark{\ref{Fermilab}},
Daniel J. Eisenstein\altaffilmark{\ref{Arizona}},
Douglas P. Finkbeiner\altaffilmark{\ref{Berkeley},\ref{Princeton},\ref{HF}},
David W. Hogg\altaffilmark{\ref{NYU}},
Robert H. Lupton\altaffilmark{\ref{Princeton}},
David J. Schlegel\altaffilmark{\ref{Princeton}}, 
Mark Subbarao\altaffilmark{\ref{Chicago}},
Neta A. Bahcall\altaffilmark{\ref{Princeton}},
John P. Blakeslee\altaffilmark{\ref{JHU}},
J. Brinkmann\altaffilmark{\ref{APO}},
Francisco J. Castander\altaffilmark{\ref{yale},\ref{chile}},
Andrew J. Connolly\altaffilmark{\ref{Pitt}}, 
Istvan Csabai\altaffilmark{\ref{Eotvos},\ref{JHU}},
Mamoru Doi\altaffilmark{\ref{Tokyo1},\ref{Tokyo2}},
Masataka Fukugita\altaffilmark{\ref{ICRR},\ref{IAS}},
Joshua Frieman\altaffilmark{\ref{Chicago},\ref{Fermilab}},
Timothy Heckman\altaffilmark{\ref{JHU}},
Gregory S. Hennessy\altaffilmark{\ref{USNO}},
\v{Z}eljko Ivezi\'{c}\altaffilmark{\ref{Princeton}},
G. R. Knapp\altaffilmark{\ref{Princeton}},
Don Q. Lamb\altaffilmark{\ref{Chicago}},
Timothy McKay\altaffilmark{\ref{UMich}},
Jeffrey A. Munn\altaffilmark{\ref{USNO}},
Robert Nichol\altaffilmark{\ref{CMU}},
Sadanori Okamura\altaffilmark{\ref{Tokyo3},\ref{Tokyo2}}, 
Donald P. Schneider\altaffilmark{\ref{PSU}},
Aniruddha R. Thakar\altaffilmark{\ref{JHU}},
and Donald G.\ York\altaffilmark{\ref{Chicago}}
}

\newcounter{address}
\setcounter{address}{1}
\altaffiltext{\theaddress}{\stepcounter{address}
University of Chicago, Astronomy \& Astrophysics
Center, 5640 S. Ellis Ave., Chicago, IL 60637\label{Chicago}}
\altaffiltext{\theaddress}{\stepcounter{address}
Department of Physics, Carnegie Mellon University, Pittsburgh, PA 15213
\label{CMU}}
\altaffiltext{\theaddress}{\stepcounter{address}
Fermi National Accelerator Laboratory, P.O. Box 500,
Batavia, IL 60510\label{Fermilab}}
\altaffiltext{\theaddress}{\stepcounter{address}
Department of Physics and Astronomy, University of Pittsburgh, Pittsburgh, PA 15620\label{Pitt}}
\altaffiltext{\theaddress}{\stepcounter{address}
Stewart Observatory, University of Arizona, 933 N. Clarry Ave., Tucson, AZ 85121\label{Arizona}}
\altaffiltext{\theaddress}{\stepcounter{address}
Department of Astronomy, University of California at Berkeley, 601 Campbell Hall, Berkeley, CA 94720\label{Berkeley}}
\altaffiltext{\theaddress}{\stepcounter{address}
Princeton University Observatory, Princeton, NJ 08544\label{Princeton}}
\altaffiltext{\theaddress}{\stepcounter{address}Hubble Fellow\label{HF}}
\altaffiltext{\theaddress}{\stepcounter{address}
Department of Physics, New York University, 4 Washington Place, New York, NY 10003\label{NYU}}
\altaffiltext{\theaddress}{\stepcounter{address}
Department of Physics \& Astronomy, The Johns Hopkins University, 3400 North Charles Street, Baltimore, MD 21218-2686\label{JHU}}
\altaffiltext{\theaddress}{\stepcounter{address}
Apache Point Observatory, 2001 Apache Point Road, P.O. Box 59, Sunspot, NM
88349-0059\label{APO}}
\altaffiltext{\theaddress}{\stepcounter{address} Yale University, P. O. Box
208101, New Haven, CT 06520\label{yale}}
\altaffiltext{\theaddress}{\stepcounter{address} Universidad de Chile, Casilla
36-D, Santiago, Chile\label{chile}}
\altaffiltext{\theaddress}{\stepcounter{address}
Department of Physics of Complex Systems, E\"otv\"os University, Budapest, H-1117 Hungary\label{Eotvos}}
\altaffiltext{\theaddress}{\stepcounter{address}
Institute of Astronomy, School of Science, University of Tokyo, Mitaka, Tokyo 181-0015, Japan\label{Tokyo1}}
\altaffiltext{\theaddress}{\stepcounter{address}
Research Center for the Early Universe, School of Science,
    University of Tokyo, Tokyo 113-0033, Japan\label{Tokyo2}}
\altaffiltext{\theaddress}{\stepcounter{address}
Institute for Cosmic Ray Research, University of Tokyo, Kashiwa 277-8582, Japan\label{ICRR}}
\altaffiltext{\theaddress}{\stepcounter{address}
Institute for Advanced Study, Olden Lane, Princeton, NJ 08540\label{IAS}}
\altaffiltext{\theaddress}{\stepcounter{address}
U.S. Naval Observatory, 3450 Massachusetts Ave., NW, Washington, DC 20392-5420\label{USNO}}
\altaffiltext{\theaddress}{\stepcounter{address}
Department of Physics, University of Michigan, 500 East University, Ann Arbor, MI 48109\label{UMich}}
\altaffiltext{\theaddress}{\stepcounter{address}
Department of Astronomy, University of Tokyo,
   Tokyo 113-0033, Japan\label{Tokyo3}}
\altaffiltext{\theaddress}{\stepcounter{address}
Department of Astronomy and Astrophysics, The Pennsylvania State University, University Park, PA 16802\label{PSU}}

%\begin{verbatim}
%bernardi@oddjob.uchicago.edu, sheth@fnal.gov,annis@fnal.gov,burles@fnal.gov,
%eisenste@cmb.as.arizona.edu,dfink@astro.princeton.edu,hogg@physics.nyu.edu,
%subbarao@oddjob.uchicago.edu,schlegel@astro.princeton.edu
%\end{verbatim}

%\maketitle

\begin{abstract}

A magnitude limited sample of nearly 9000 early-type galaxies, 
in the redshift range $0.01 \le z \le 0.3$, was selected from the 
Sloan Digital Sky Survey  using morphological and spectral criteria.  
The sample was used to study how early-type galaxy observables, 
including luminosity $L$, effective radius $R_o$, surface brightness 
$I_o$, color, and velocity dispersion $\sigma$, are correlated with 
one another.  Measurement biases are understood with mock catalogs 
which reproduce all of the observed scaling relations and their 
dependences on fitting technique.  
At any given redshift, the intrinsic distribution of luminosities, 
sizes and velocity dispersions in our sample are all approximately 
Gaussian.  A maximum likelihood analysis shows that 
$\sigma\propto L^{0.25\pm 0.012}$, $R_o\propto L^{0.63\pm 0.025}$, 
and $R_o\propto I^{-0.75\pm 0.02}$ in the $r^*$ band.  In addition, 
the mass-to-light ratio within the effective radius scales as 
$M_o/L \propto L^{0.14\pm 0.02}$ or $M_o/L\propto M_o^{0.22\pm 0.05}$, 
and galaxies with larger effective masses have smaller effective 
densities:  $\Delta_o\propto M_o^{-0.52\pm 0.03}$.   
These relations are approximately the same in the $g^*$, $i^*$ and 
$z^*$ bands.  
Relative to the population at the median redshift in the sample, 
galaxies at lower and higher redshifts have evolved only little, 
with more evolution in the bluer bands.  
The luminosity function is consistent with weak passive luminosity 
evolution and a formation time of about 9~Gyrs ago.  

\end{abstract}  

\section{Introduction}
This is the second of four papers in which the properties of 
$\sim 9000$ early-type galaxies, in the redshift range 
$0.01\le z\le 0.3$ are studied.  
Paper I (Bernardi et al. 2003a) describes how the sample was selected 
from the SDSS database.  The sample is essentially magnitude limited, 
and the galaxies in it span a wide range of environments.  
Each galaxy in the sample has measured values of luminosity $L$, 
effective radius $R_o$ and surface brightness $I_o=(L/2)/R_o^2$ in 
four bands ($g^*$, $r^*$, $i^*$ and $z^*$), a velocity dispersion 
$\sigma$, a redshift, and an estimate of the local density.  

Section~\ref{lf} of the present paper shows that the luminosity function 
of the galaxies in our sample, when expressed as a function of absolute 
magnitude, is well described by a Gaussian form, and that the 
luminosities in the population as a whole appear to be evolving passively.  
Section~\ref{lx} studies the distribution of (the logarithm of) velocity 
dispersion, size, surface-brightness, effective mass and effective density 
at fixed luminosity; all of these are quite well described by Gaussian 
forms, suggesting that the intrinsic distributions of log(size) and 
log(velocity dispersion) are, like the distribution of log(luminosity), 
approximately Gaussian.  
Maximum-likelihood estimates of these and other correlations, which 
include the Faber-Jackson relation, the mass-to-light ratio, 
the Kormendy relation and a mass--density relation are presented in 
Section~\ref{ML3d}.  
Appendix~\ref{simul} describes a method for generating accurate mock 
complete and magnitude-limited galaxy catalogs, which are useful for 
assessing the relative importance of evolution and selection effects.  
The procedure used to estimate errors on our results is discussed in 
Appendix~\ref{compcat}.  

Paper~III (Bernardi et al. 2003b) of this series places special emphasis 
on the Fundamental Plane relation between size, surface brightness and 
velocity dispersion.  It shows how the FP depends on waveband, color, 
redshift and environment.  Paper~IV (Bernardi et al. 2003c) uses the 
colors and spectra of these galaxies to provide information about the 
chemical evolution of the early-type population.  

Except where stated otherwise, we write the Hubble constant as 
$H_0=100\,h~\mathrm{km\,s^{-1}\,Mpc^{-1}}$, and we perform our 
analysis in a cosmological world model with 
$(\Omega_{\rm M},\Omega_{\Lambda},h)=(0.3,0.7,0.7)$, where 
$\Omega_{\rm M}$ and $\Omega_{\Lambda}$ are the present-day scaled 
densities of matter and cosmological constant.  
In such a model, the age of the Universe at the present time is 
$t_0=9.43h^{-1}$ Gyr.  For comparison, an Einstein-de Sitter model has 
$(\Omega_{\rm M},\Omega_{\Lambda})=(1,0)$ and $t_0=6.52h^{-1}$~Gyr.  
We frequently use the notation $h_{70}$ as a reminder that we have 
set $h=0.7$.  Also, we will frequently be interested in the 
logarithms of physical quantities.  Our convention is to set 
$R\equiv\log_{10}R_o$ and $V\equiv \log_{10}\sigma$, where $R_o$ 
and $\sigma$ are effective radii in $~h_{70}^{-1}$~kpc and velocity 
dispersions in km~s$^{-1}$, respectively.  

\section{The luminosity function}\label{lf}
Our sample is magnitude limited (Table~1 of Paper~I gives the 
magnitude limits in the different bands).  Therefore, we measure 
the luminosity function of the galaxies in our sample using 
two techniques.  The first uses volume limited catalogs, and the 
second uses a maximum likelihood procedure 
(Sandage, Tammann \& Yahil 1979; Efstathiou, Ellis \& Peterson 1988).  

In the first method, we divide our parent catalog into many volume 
limited subsamples; this was possible because the parent catalog is so 
large.  When doing this, we must decide what size volumes to choose.  
We would like our volumes to be as large as possible so that each volume 
represents a fair sample of the Universe.  On the other hand, the volumes 
must not be so large that evolution effects are important.  In addition, 
because our catalog is cut at the bright as well as the faint end, 
large-volume subsamples span only a small range in luminosities.  
Therefore, we are forced to compromise:  we have chosen to make the 
volumes about $\Delta z=0.04$ thick, because 
$c\Delta z/H\approx 120h^{-1}$Mpc is larger than the largest 
structures seen in numerical simulations of the cold dark matter family 
of models (e.g., Colberg et al. 2000).  The catalogs are extracted from 
regions which cover a very wide angle on the sky, so the actual volume of 
any given volume limited catalog is considerably larger 
than ($120h^{-1}$Mpc)$^3$.  Therefore, this choice should provide volumes 
which are large enough in at least two of the three coordinate directions 
that they represent fair samples, but not so large in the redshift 
direction that the range in luminosities in any given catalog is small, 
or that evolution effects are washed out.  

The volume-limited subamples are constructed as follows.  
First, we specify the boundaries in redshift of the catalog:  
$z_{\rm min}$ and $z_{\rm max}=z_{\rm min} + 0.04$.  
In the context of a world model, these redshift limits, 
when combined with the angular size of the catalog, can be used to 
compute a volume.  This volume depends on $z_{\rm min}, z_{\rm max}$ 
and the world model:  as our fiducial model we set $\Omega_M = 0.3$ 
and $\Omega_\Lambda=1-\Omega_M$.  (Our results hardly change if we 
use an Einstein de-Sitter model instead.)  We then compute the 
K-corrected limiting luminosities $L_{\rm max}(z_{\rm min})$ and 
$L_{\rm min}(z_{\rm max})$ given the apparent magnitude limits, 
the redshift limits, and the assumed cosmology.  
A galaxy $i$ is included in the volume limited subsample if 
$z_{\rm min}\le z_i\le z_{\rm max}$ and 
$L_{\rm min}\le L_i\le L_{\rm max}$.  The luminosity function for the 
volume limited subsample is obtained by counting the number of galaxies 
in a luminosity bin and dividing by the volume of the subsample.  

\begin{figure}
\centering
\epsfxsize=\hsize\epsffile{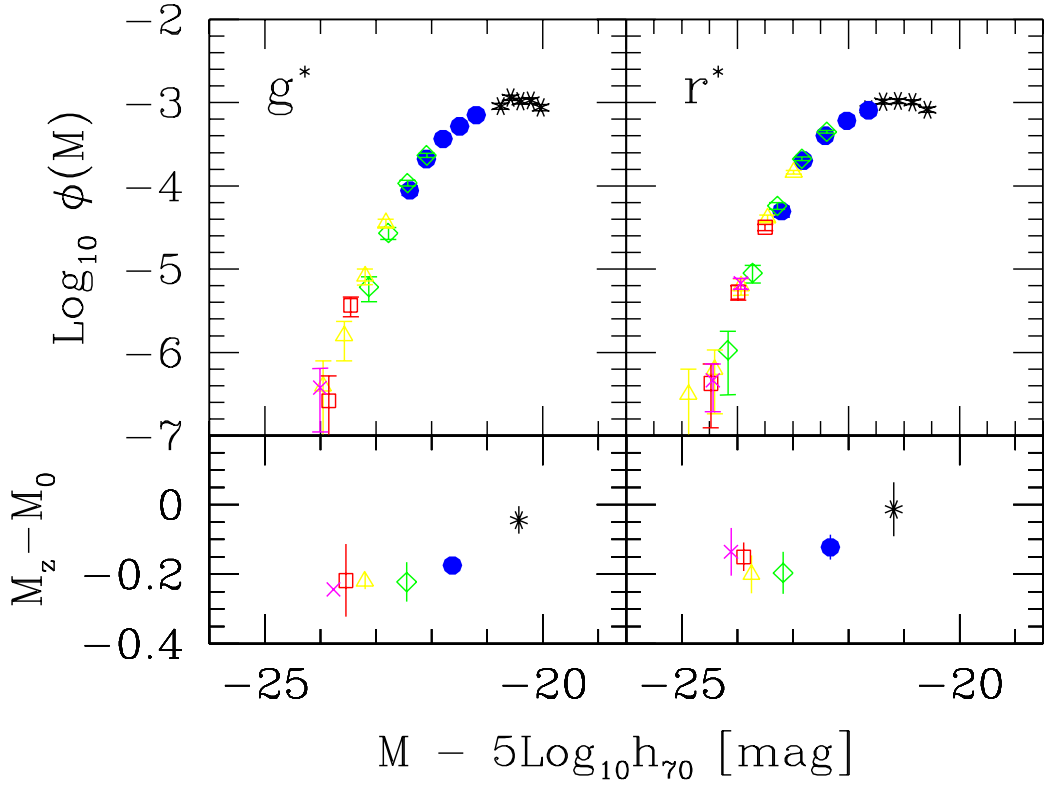}
\vspace{-3cm}
\caption{Luminosity functions in the $g^*$ and $r^*$ bands.  Stars, 
circles, diamonds, triangles, squares and crosses show measurements in 
volume limited catalogs which are adjacent in redshift of width 
$\Delta z=0.04$, starting from a minimum of $z_{\rm min}=0.04$.  
Top panels show that the higher redshift catalogs contribute at the 
bright end only.  At the same comoving density, the symbols which 
represent the higher redshift catalogs tend to be displaced slightly 
to the left of the those which represent the lower redshift catalogs.  
Bottom panels show this small mean shift towards increasing luminosity 
with increasing redshift.}  
\label{lfev}
\end{figure}

The top panels in Figure~\ref{lfev} show the result of doing this in the 
$g^*$ and $r^*$ bands.  Stars, circles, diamonds, triangles, squares and 
crosses show measurements in volume limited catalogs which have 
$z_{\rm min}=0.04$, 0.08, 0.12, 0.16, 0.20, and 0.24 and 
$z_{\rm max} = z_{\rm min}+0.04$.  Each subsample contains more than 
five hundred galaxies, except for the two most distant, which each 
contain about one hundred.  As one would expect, the nearby volumes 
provide the faint end of $\phi(M)$, and the more distant volumes show 
the bright end.  The extent to which the different volume limited 
catalogs all trace out the same curve is a measure of how little the 
luminosity function at low and high redshifts differs from that at the 
median redshift.  

The bottom panels in Figure~\ref{lfev} show evidence that, in fact, the 
galaxies in our data show evidence for a small amount of evolution:  at 
fixed comoving density, the higher redshift population is slightly brighter 
than that at lower redshifts.  Although volume-limited catalogs provide 
model-independent measures of this evolution, the test is most sensitive 
when a large range of luminosities can be probed at two different 
redshifts.  Because the SDSS catalogs are cut at both the faint and the 
bright ends, our test for evolution is severely limited.  Nevertheless, 
the small trends we see are both statistically significant, and 
qualitatively consistent with what one expects of a passively evolving 
population.  (Note that our sample contains only early-type galaxies.  
Blanton et al. 2001 study the luminosity function in an SDSS sample which 
contains all galaxy types, but they ignore evolution effects.  Since 
late-type galaxies are expected to evolve more rapidly than early-types, 
it is important to redo Blanton et al.'s analysis after allowing for 
evolution.)  

Before we make more quantitative conclusions, notice that a bell-like 
Gaussian shape would provide a reasonable description of the luminosity 
function.  Although early-type galaxies are expected to have red colors, 
our sample was not selected using any color information.  It is reassuring, 
therefore, that the Gaussian shape we find here also provides a good fit to 
the luminosity function of the redder objects in the SDSS parent catalog 
(see the curves for the two reddest galaxy bins in Fig.14 of 
Blanton et al. 2001).  A Gaussian form also provides a reasonable 
description of the luminosity function of early-type galaxies in the 
CNOC2 survey (Lin et al. 1999, even though they actually fit a Schechter 
function to their measurements).  The 2dFGRS galaxies classified as being 
of Type 1 by Madgwick et al. (2002) should be similar to early-types.  
Their Type 1's extend to considerably fainter absolute magnitudes than 
our sample and the shape of the luminosity function they report is quite 
different from ours.  This is probably because the population of 
early-type galaxies at faint absolute magnitudes is quite different 
from the brighter ones (e.g., Sandage \& Perelmuter 1990).  In any case, 
their Schechter function fits underestimate the number density of 
luminous Type 1 galaxies---a Gaussian tail would provide a significantly 
better fit.  

Given that the Gaussian form provides a good description of our data, 
we use the maximum-likelihood method outlined by 
Sandage, Tammann \& Yahil (1979) to estimate the parameters of the 
best-fitting luminosity function.  For magnitude limited samples which 
are small and shallow, this is the method of choice.  For a sample such 
as ours, which spans a sufficiently wide range in redshifts that evolution 
effects might be important, the method requires a model for the evolution.  
We parametrize the luminosity evolution similarly to Lin et al. (1999).  
That is to say, if we were solving only for the luminosity function, 
then the likelihood function we maximize would be  
\begin{eqnarray}
{\cal L} &=& \prod_i{\phi(M_i,z_i|Q,M_*,\sigma_M)\over S(z_i|Q,M_*,\sigma_M)},
 \qquad {\rm where} \nonumber\\
\phi(M_i,z_i|Q,M_*,\sigma_M) &=& {\phi_*\over\sqrt{2\pi\sigma_M^2}}\,
           \exp\left(-{[M_i-M_*+Qz_i]^2\over 2\sigma_M^2}\right), \nonumber\\
S(z_i|Q,M_*,\sigma_M) &=& 
 \int_{M_{\rm min}(z_i)}^{M_{\rm max}(z_i)} dM\,\phi(M,z_i|Q,M_*,\sigma_M),
\label{mlphi}
\end{eqnarray}
$M_{\rm min}(z_i)$ and $M_{\rm max}(z_i)$ denote the minimum and maximum 
absolute magnitudes at $z_i$ which satisfy the apparent magnitude limits 
of the survey, and $i$ runs over all the galaxies in the catalog.  
(At small $z$, this parametrization of the evolution in absolute magnitude 
implies that the luminosity evolves as $L_*(z)/L_*(0) \approx (1+z)^q$, 
with $q = Q\,{\rm ln}(10)/2.5$.  Note that, in assuming that only $M_*$ 
evolves, this model assumes that there is no differential evolution in 
luminosities, i.e., that luminous and not so luminous galaxies evolve 
similarly.    

\begin{figure}
\centering
\epsfxsize=\hsize\epsffile{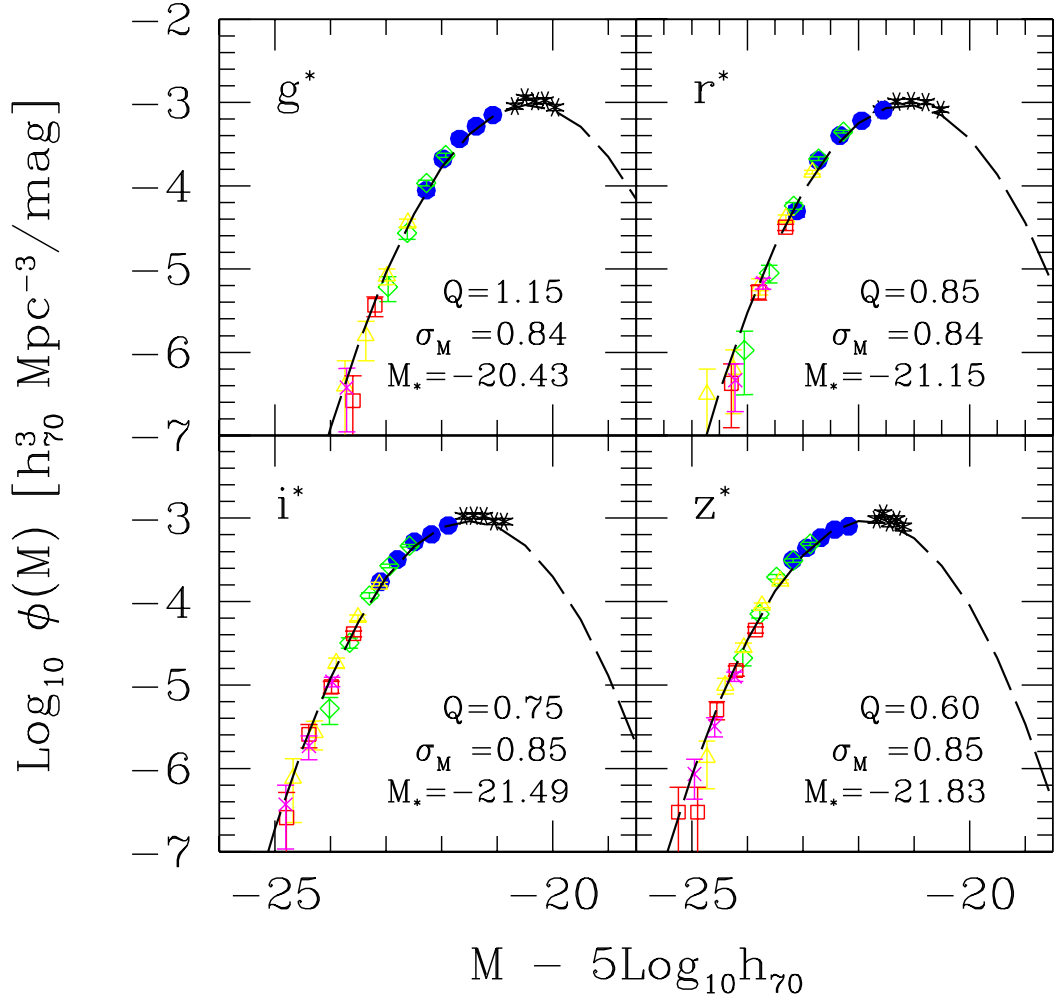}
\vspace{-1cm}
\caption{Luminosity functions in the $g^*$, $r^*$, $i^*$ and $z^*$ 
bands, corrected for pure luminosity evolution.  
Symbols with error bars show the estimates from our various 
volume limited catalogs; the higher redshift catalogs contribute at 
the bright end only.  Dashed curves show the shape of the Gaussian 
shaped luminosity function which maximizes the likelihood of seeing 
this data at redshift $z=0$.}  
\label{lf4}
\end{figure}

Figure~\ref{lf4} shows the result of estimating the luminosity function 
in this way in the $g^*$, $r^*$, $i^*$ and $z^*$ bands.  Later in this 
paper, we will solve simultaneously for the joint distribution of 
luminosity, size and velocity dispersion; it is the parameters which 
describe the luminosity function of this joint solution which are shown 
in Fig.~\ref{lf4}.  The dashed lines in each panel show the Gaussian shape 
of the luminosity function at redshift $z=0$.  For comparison, the symbols 
show the measurements in the same volume limited catalogs as before, 
except that now  we have subtracted the maximum likelihood estimate of 
the luminosity evolution from the absolute magnitudes $M$ before plotting 
them.  If the model for the evolution is accurate, then the different 
symbols should all trace out the same smooth dashed curve.  

\begin{figure}
\centering
\epsfxsize=\hsize\epsffile{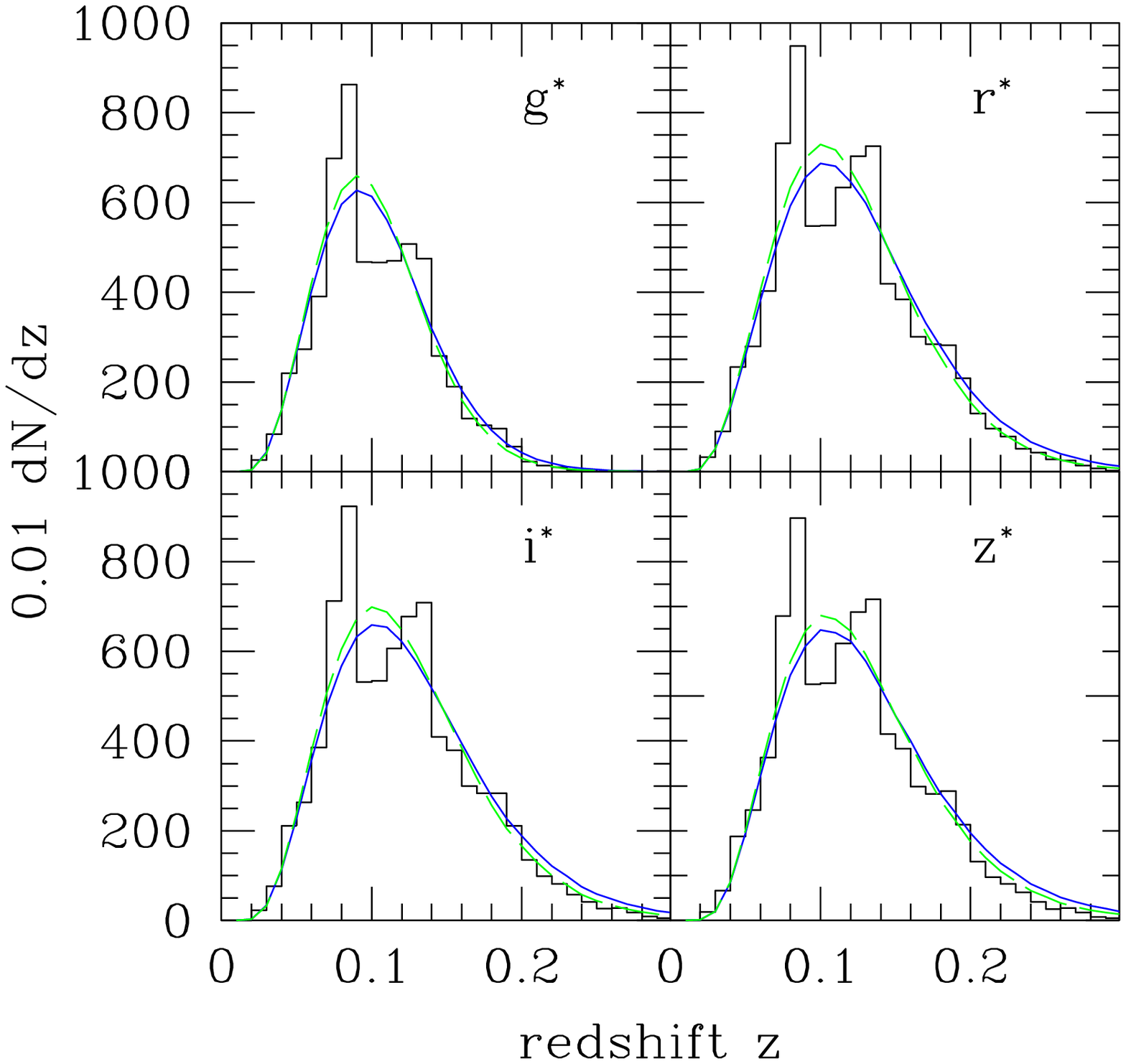}
\vspace{-0.5cm}
\caption{The number of galaxies as a function of redshift in our sample.  
Solid curves show the predicted counts if the comoving number densities 
are constant, but the luminosities brighten systematically with redshift: 
$M_*(z)= M_*(0)-Qz$ with $Q$ given by the previous figure.
Dashed curves show what one predicts if there is no evolution whatsoever, 
and the luminosity function is fixed to the value it has at the median 
redshift of our sample ($z=0.1$).
%; or b) there is density as well as luminosity evolution, as reported by 
%Lin et al. (1999); or 
%c) the comoving number densities do not change, but the more luminous 
%galaxies evolve less rapidly than less luminous galaxies.
}
\label{nz4}
\end{figure}

The comoving number density of the galaxies in this sample is 
$\phi_* = 5.8\pm 0.3\times 10^{-3}h^3$Mpc$^{-3}$ in all four bands.  
Because the different bands have different apparent magnitude 
limits, and they were fit independently of each other, it is reassuring 
that the same value of $\phi_*$ works for all the bands.  
For similar reasons, it is reassuring that the best-fit values of $M_*$ 
imply rest-frame colors at $z=0$ of $g^*-r^* = 0.72$, $r^*-i^* = 0.34$, 
and $r^*-z^* = 0.68$, which are close to those of the models which we used 
to compute our K-corrections (Appendix~A of Paper I), even though no a 
priori constraint was imposed on what these rest-frame colors should be.  

The histograms in each of the four panels of Figure~\ref{nz4} show the 
number of galaxies observed as a function of redshift in the four bands.  
The peak in the number counts at $z\sim 0.08$ is also present in the 
full SDSS sample, which includes late-types, and, perhaps more 
surprisingly, an overdensity at this same redshift is also present 
in the 2dF Galaxy Redshift Survey.  (The second bump at $z\sim 0.13$ 
is also present in the 2dFGRS counts.)    
The solid curves show what we expect to see for the evolving Gaussian 
function fits---the curves provide a reasonably good fit to the observed 
counts, although they slightly overestimate the numbers at high redshift 
in the redder wavebands.  
For comparison, the dashed curves show what is expected if the luminosities 
do not evolve and the no-evolution luminosity function is given by the one 
at the median redshift (i.e., a Gaussian with mean $M_*-0.1Q$).  
Although the fit to the high-redshift tail is slightly better, this 
no evolution model cannot explain the trends shown in the bottom panel 
of Figure~\ref{lfev}.  
Moreover, a Bruzual \& Charlot (2003) passive evolution model with a 
formation time of 9~Gyrs ago, predicts that the rest-frame luminosities 
at redshift $z=0.2$ should be brighter than those at $z=0$ by 
0.3, 0.26, 0.24, and 0.21~mags in $g^*$, $r^*$, $i^*$ and $z^*$ 
respectively---not far off from what we estimate.  

The bottom panels in Figure~\ref{lfev} suggest two possible reasons why 
our model of pure luminosity evolution overestimates $dN/dz$ at higher 
$z$.  One possibility is that the comoving number densities are 
decreasing slightly with redshift.  A small amount of density evolution 
is not unexpected, because early-type galaxy morphologies may evolve 
(van Dokkum \& Franx 2001), and our sample is selected on the basis of 
a fixed morphology.  If we allow a small amount of density as well 
as luminosity evolution, and we use $\phi_*(z) = 10^{0.4Pz}\phi_*(0)$ 
with $P\approx -2$, as suggested by the results of Lin et al. (1999), 
then the resulting $dN/dz$ curves are also well fit by the dashed curves. 
A second possibility follows from the fact that we only observe the 
most luminous part of the higher redshift population.  If the most 
luminous galaxies at any given time are also the oldest, then one might 
expect the bright end of the luminosity function to evolve less rapidly 
than the fainter end.  
The curvature seen in the bottom panel of Figure~\ref{lfev} suggests that 
although the evolution of the fainter objects in our sample (which we only 
see out to low redshifts) is consistent with formation times of 9~Gyrs 
ago, the brighter objects are not.  Models of differential evolution 
in the luminosities also predict $dN/dz$ distributions which are in 
better agreement with the observed counts at high redshift.  
Since the evolution of the luminosity function is small, we prefer to 
wait until we are able to make more accurate K-corrections before 
accounting for either of these other possibilities more carefully.  
Therefore, in what follows, we will continue to use the model with 
pure luminosity evolution.  

Repeating the exercise described above but for an Einstein--de-Sitter 
model yields qualitatively similar results, although the actual values 
of $M_*$ and $\phi_*$ are slightly different.  At face value, the fact 
that we see so little evolution in the luminosities argues for a 
relatively high formation redshift: the Bruzual \& Charlot (2003) 
models indicate that $t_{\rm form}\sim 9$~Gyrs.  

\section{Observed correlations: Distributions at fixed luminosity}\label{lx}
This Section presents scatter plots between different observables 
$X$ and luminosity.  This is done because, except for a cut at small 
velocity dispersions, our sample was selected by luminosity alone.  
This means that the distributions of $X$ at fixed luminosity are not 
biased by the selection cut (e.g., Schechter 1980).  
The distribution of $X$ at fixed $L$ is shown to be reasonably well 
described by a Gaussian for all the choices of $X$ we consider.  
This simplifies the maximum likelihood analysis described in 
Section~\ref{ML3d} which we use to estimate a number of observed 
correlations (it is also used in Paper~III to estimate the parameters 
of the Fundamental Plane).

The best way to think of any absolute magnitude $M$ versus $X$ 
scatter plot is to imagine that, at fixed absolute magnitude $M$, 
there is a distribution of $X$ values.  The scatter plot then shows 
the joint distribution 
\begin{equation}
\phi(M,X|z)\,{\rm d}M\,{\rm d}X = {\rm d}M\,\phi(M|z)\ p(X|M,z){\rm d}X ,
\label{plx}
\end{equation}
where $\phi(M,X|z)$ denotes the density of galaxies with $X$ and $M$ at 
$z$, and $\phi(M|z)$ is the luminosity function at $z$ which we computed 
in Section~\ref{lf}.  One of the results of this section is to show 
that the shape of $p(X|M,z)$ is simple for most of the relations of 
interest.  

The mean value of $X$ at fixed $M$ is independent of the fact that our 
catalogs are magnitude limited.  Therefore, we estimate the parameters 
of linear relations of the form:  
\begin{equation}
(X-X_*) = {-0.4\,(M-M_*)\over S},
\label{xlregress}
\end{equation}
where $M=-2.5\log_{10}L$ is the absolute magnitude 
and $X$ is the observable (for example, we will study 
$X=\log_{10}\sigma$, $\log_{10}R_o$ or $\mu_o=-2.5\log_{10}I_o$).  
For each volume limited catalog, we fit for the slope $S$ and 
zero-point of the linear relation.  If there really were a 
linear relation between $M$ and $X$, and neither $X$ nor $M$ 
evolved, then the slopes and zero-points of the different 
volume limited catalogs would be the same.  

\begin{figure}
\centering
\epsfxsize=\hsize\epsffile{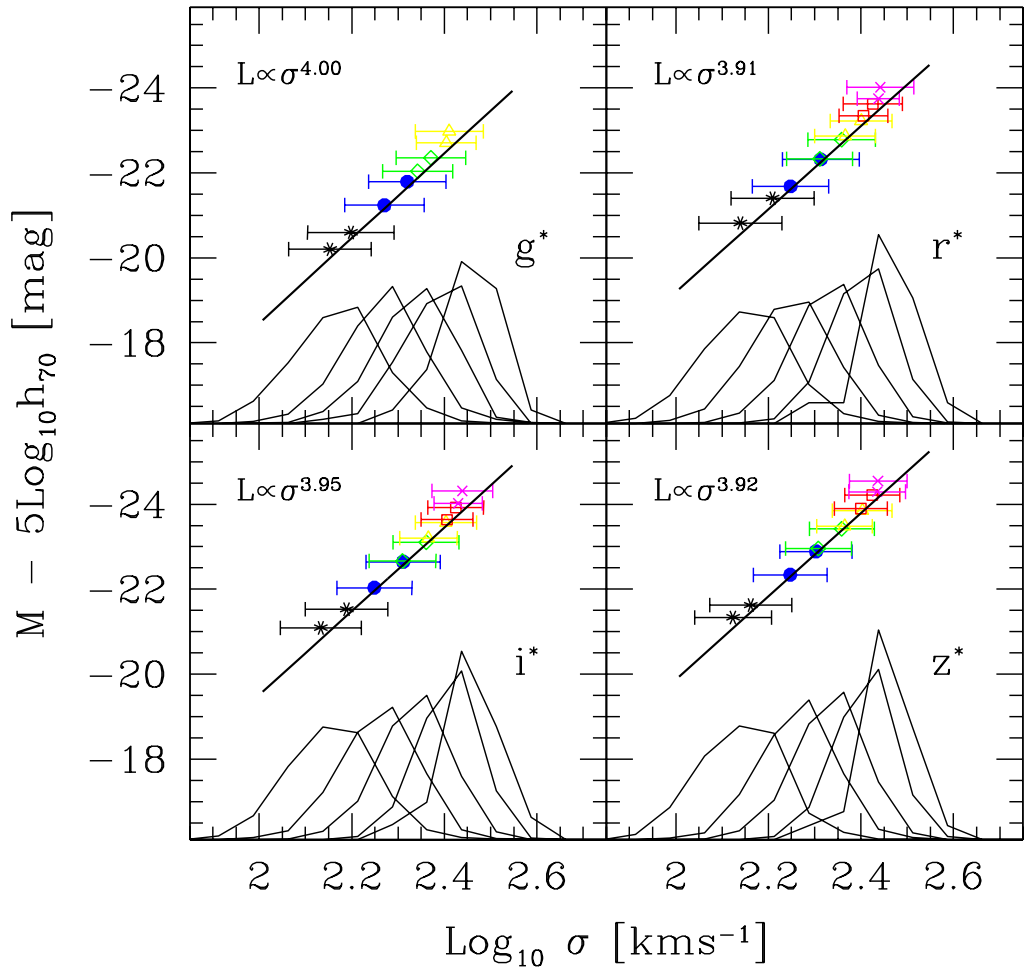}
\vspace{-1.cm}
\caption{Relation between luminosity $L$ and velocity dispersion $\sigma$.  
Stars, circles, diamonds, triangles, squares and crosses show the 
error-weighted mean value of $\log_{10}\sigma$ for a small range in 
luminosity in each volume limited catalog (see text for details).  
(Only catalogs containing more than one hundred galaxies are shown.)  
Error bars show the rms scatter around this mean value.  
Solid line shows the maximum-likelihood estimate of this relation, 
and the label in the top left shows the scaling it implies.  
Histograms show the distribution of 
$\log_{10}\sigma$ in small bins in luminosity.  They were obtained 
by stacking together non-overlapping volume limited catalogs to construct 
a composite catalog, and then dividing the composite catalog into five 
equal size bins in luminosity.}
\label{ls}
\end{figure}

To illustrate, the different symbols in Figure~\ref{ls} show 
$\langle\log_{10}\sigma|M\rangle$, the Faber--Jackson relation 
(Faber \& Jackson 1976), in our dataset.  
Most datasets in the literature are consistent with the scaling 
$\langle\sigma|L\rangle\propto L^{1/4}$, approximately independent 
of waveband.  For example, Forbes \& Ponman (1999), using a compilation 
of data from Prugniel \& Simien (1996) report $L\propto\sigma^{3.92}$ 
in the B-band.  At longer wavelengths Pahre et al. (1998) report 
$L_K\propto\sigma^{4.14\pm 0.22}$ in the K-band, with a scatter of 
0.93 mag.  

Stars, circles, diamonds, triangles, squares and crosses show the 
relation measured in volume limited catalogs of successively higher 
redshift (redshift limits are the same as in Figure~\ref{lfev}).  
The galaxies in each subsample were further divided into two equal-sized 
parts based on luminosity.  The symbols with error bars show the mean 
$\log_{10}\sigma$ for each of these small bins in $M$, and the rms 
spread around it (note that the error on the mean is smaller than the 
size of the symbols in all but the highest redshift catalogs).  The 
solid line shows the maximum-likelihood estimate of the slope of this 
relation at $z=0$, which we describe in Section~\ref{prjct}.  
Comparison with this line shows that the higher redshift population 
is slightly brighter.  The slope of this line is shown in the top of 
each panel:  $\sigma\propto L^{1/4}$, approximately, in all the bands, 
consistent with the literature.  The zero point, however, is different; 
at fixed luminosity, the objects in our sample have velocity dispersions 
which are smaller than those reported in the literature by about 
$log_{10}\sigma = 0.05$.  

\begin{figure}
\centering
\epsfxsize=\hsize\epsffile{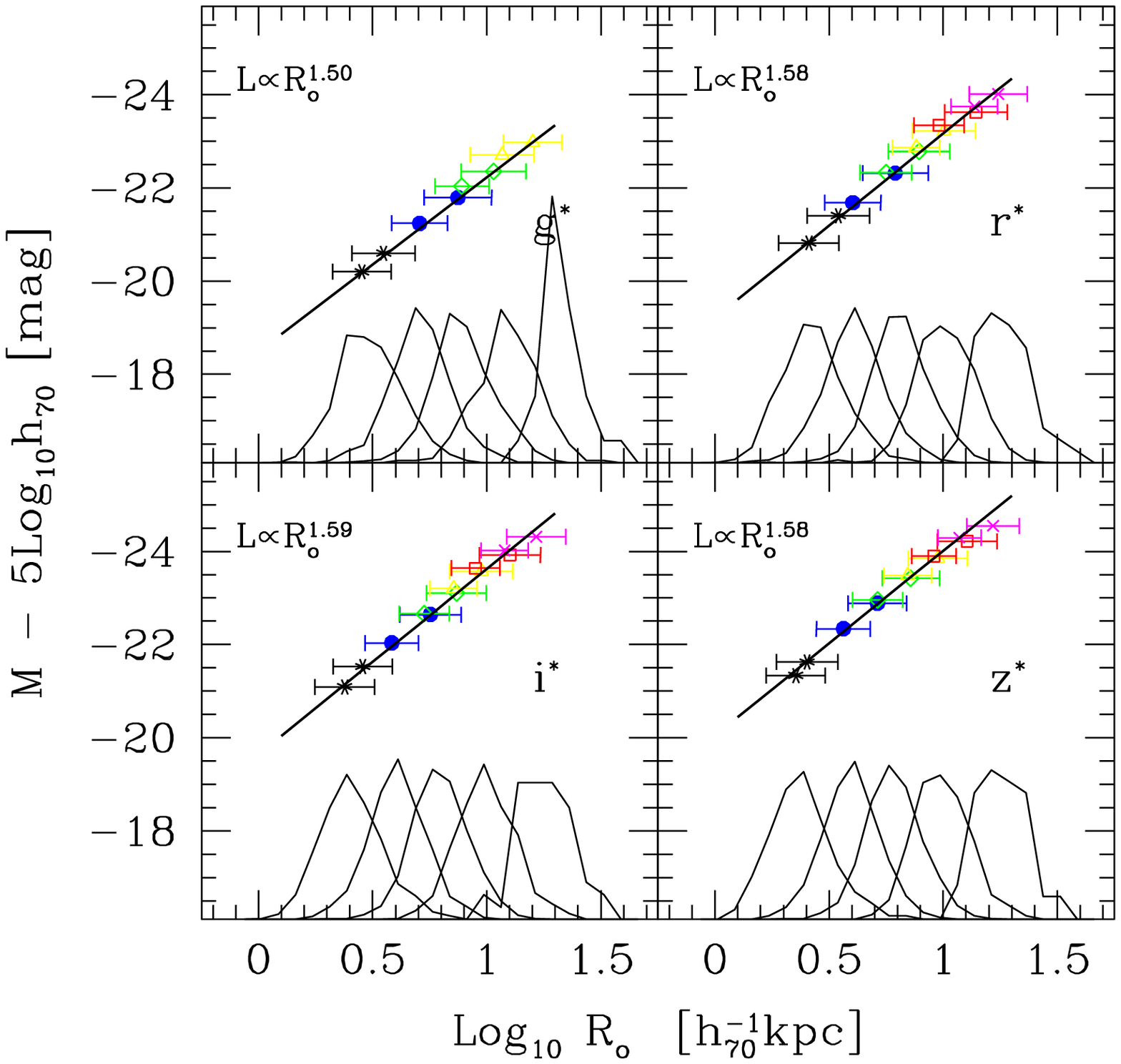}
\vspace{-1.cm}
\caption{Same as previous figure, but for the relation between 
luminosity $L$ and effective radius $R_o$.}
\label{lre}
\end{figure}

We have enough data that we can actually do more than simply measure the 
mean $X$ at fixed $M$; we can also compute the distribution around the 
mean.  If we do this for each catalog, then we obtain distributions 
which are approximately Gaussian in shape, with dispersions which 
depend on the range of luminosities which are in the subsample.  Rather 
than showing these, we created a composite catalog by stacking together 
the galaxies from the nonoverlapping volume limited catalogs, and we 
then divided the composite catalog into five equal sized bins in 
luminosity.  The histograms in the bottom of the plot show the shapes 
of the distribution of velocities in the different luminosity bins.  
Except for the lowest and highest redshift catalogs for which the 
statistics are poorest, the different distributions have almost the 
same shape; only the mean changes.  

One might have worried that the similarity of the distributions is a 
signature that they are dominated by measurement error.  This is not 
the case:  the typical measurement error is about a factor of two smaller 
than the rms of any of these distributions.  If we assume that the 
measurement errors are Gaussian-distributed, then the distributions we 
see should be the true distribution broadened by the Gaussian from the 
measurement errors.  The fact that the observed distributions are well 
approximated by Gaussians suggests that the true intrinsic distributions 
are also Gaussian.  The fact that the width of the intrinsic distribution 
is approximately independent of $M$ considerably simplifies the maximum 
likelihood analysis presented in the next section.  

It is well known that color is strongly correlated with velocity 
dispersion (Paper~IV of this series shows the color$-\sigma$ relation 
in our sample).  One consequence of this is that residuals from the 
$\sigma-L$ relation shown in Figure~\ref{ls} correlate strongly with 
color:  at fixed magnitude, the redder galaxies have the highest velocity 
dispersions.  In addition, as a whole, the reddest galaxies populate 
the high $\sigma$ part of the relation.  
Forbes \& Ponman (1999) reported that residuals from the Faber--Jackson 
relation correlate with age.  If color is an indicator of age and/or 
metallicity, then our finding is qualitatively consistent with theirs:  
the typical age/metallicity varies along the Faber--Jackson relation.  

A similar study of the relation between the luminosities and sizes 
of galaxies is shown in Figure~\ref{lre}.  
Schade et al. (1997) find $L_B\propto R_o^{4/3}$ in the B band, 
whereas, at longer wavelengths, Pahre et al. (1998) find 
$L_K\propto R_o^{7/4}$ with an rms of 0.88~mag.  This 
suggests that the relation depends on wavelength.  
We find $\langle R_o|L\rangle \propto L^{2/3}$ in $g^*$, but 
$\langle R_o|L\rangle \propto L^{3/5}$ in the other bands.  
The distribution $p(\log_{10}R_o|M)$ is also reasonably well fit 
by a Gaussian, with a mean which increases with luminosity, and a 
dispersion which is approximately independent of $M$.  
The rms around the mean is about one and a half times larger than 
the rms around the mean $\sigma-L$ relation.  
We argue in Paper~IV that the color--magnitude and color--size 
relations are a consequence of the color$-\sigma$ correlation.  
If this is correct, then residuals from the $R_o-L$ relation, should not 
correlate with size or magnitude.  We have checked that this is correct, 
although we have not included a plot showing this explicitly.  

\begin{figure}
\centering
\epsfxsize=\hsize\epsffile{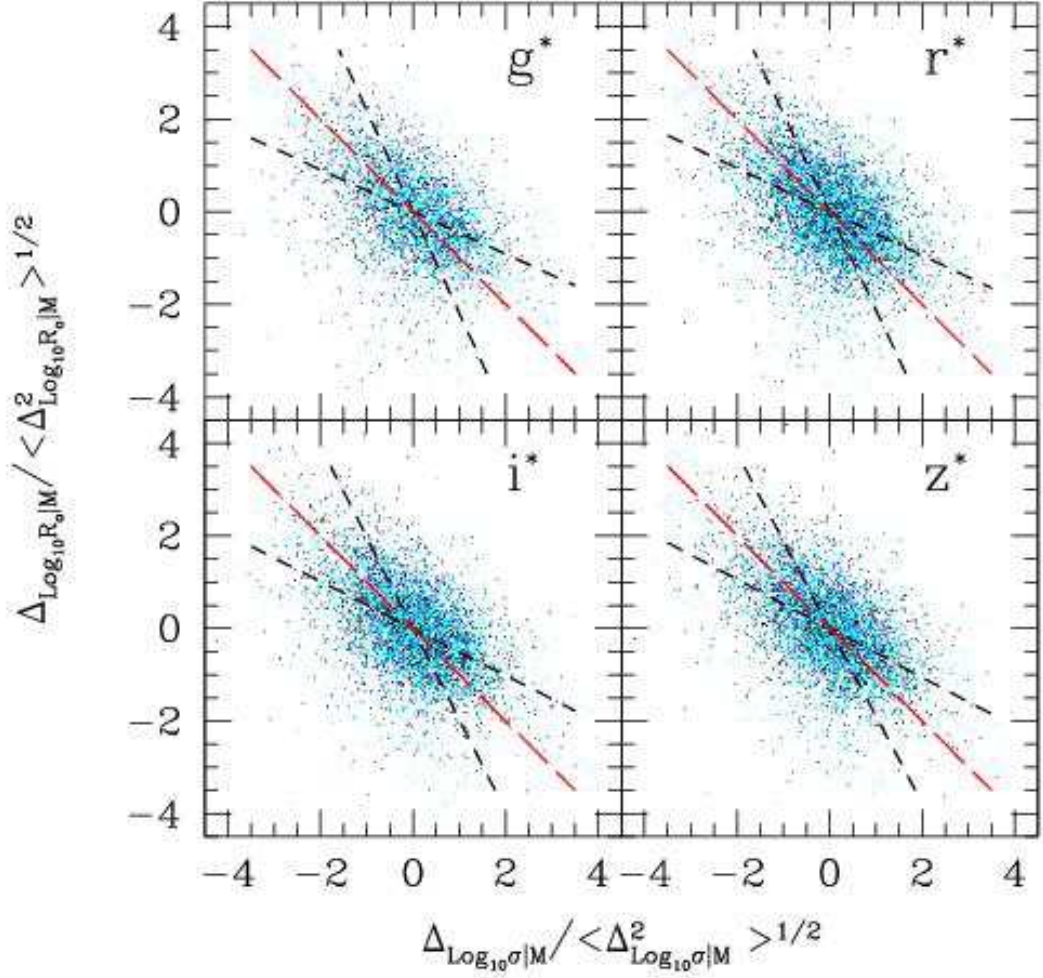}
\caption{Residuals of the $R_o-L$ relation are anti-correlated with 
residuals of the $\sigma-L$ relation; galaxies of the same luminosity 
which are smaller than expected have larger velocity dispersions than 
expected.  Plot shows the residuals normalized by their rms value.  
Short-dashed lines show forward and inverse fits to the scatter plots, 
and long-dashed line in between the other two shows  
$\Delta_{R_o|M}/\sigma_{R|M} = -\Delta_{\sigma-M}/\sigma_{V|M}$}.  
\label{resid}
\end{figure}

\begin{figure}
\centering
\epsfxsize=\hsize\epsffile{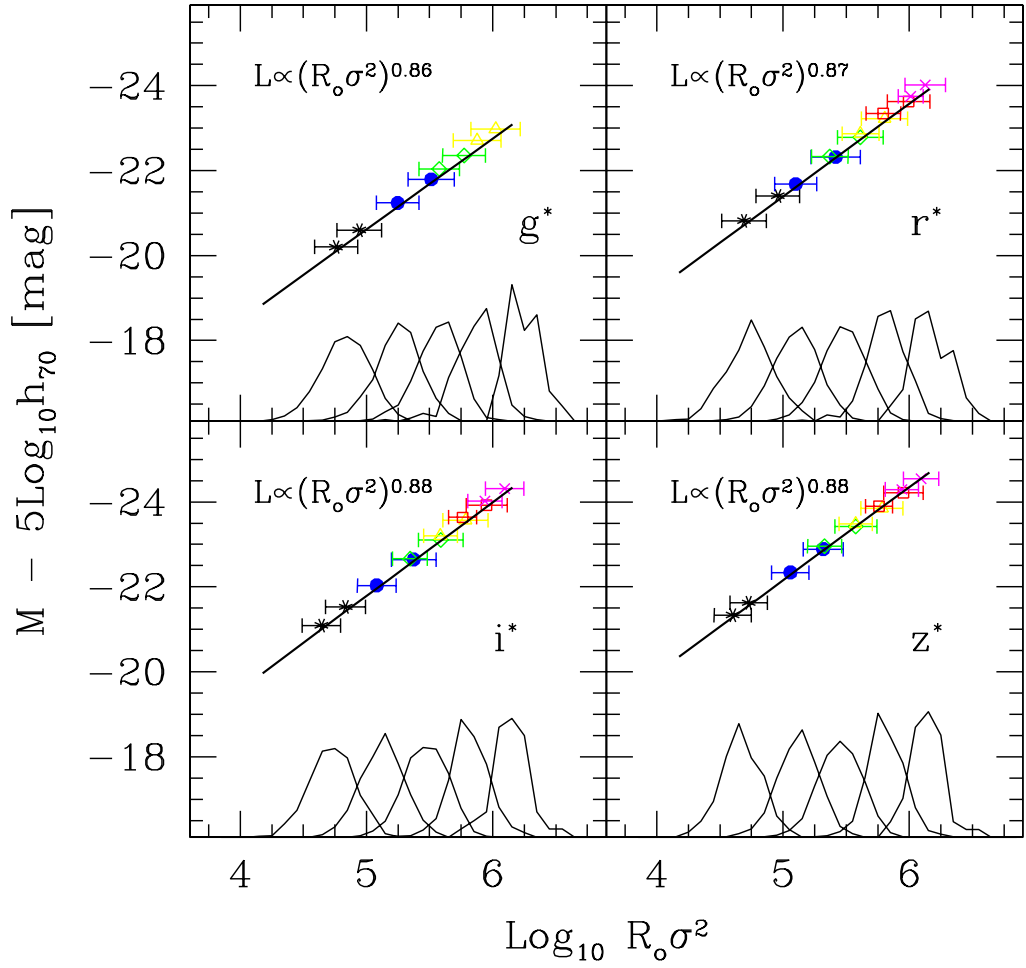}
\vspace{-1.cm}
\caption{Same as Figure~\ref{ls}, but for the relation between 
luminosity $L$ and the combination $R_o\sigma^2$, which is supposed 
to be a measure of mass.}
\label{lmass}
\end{figure}

\begin{figure}
\centering
\epsfxsize=\hsize\epsffile{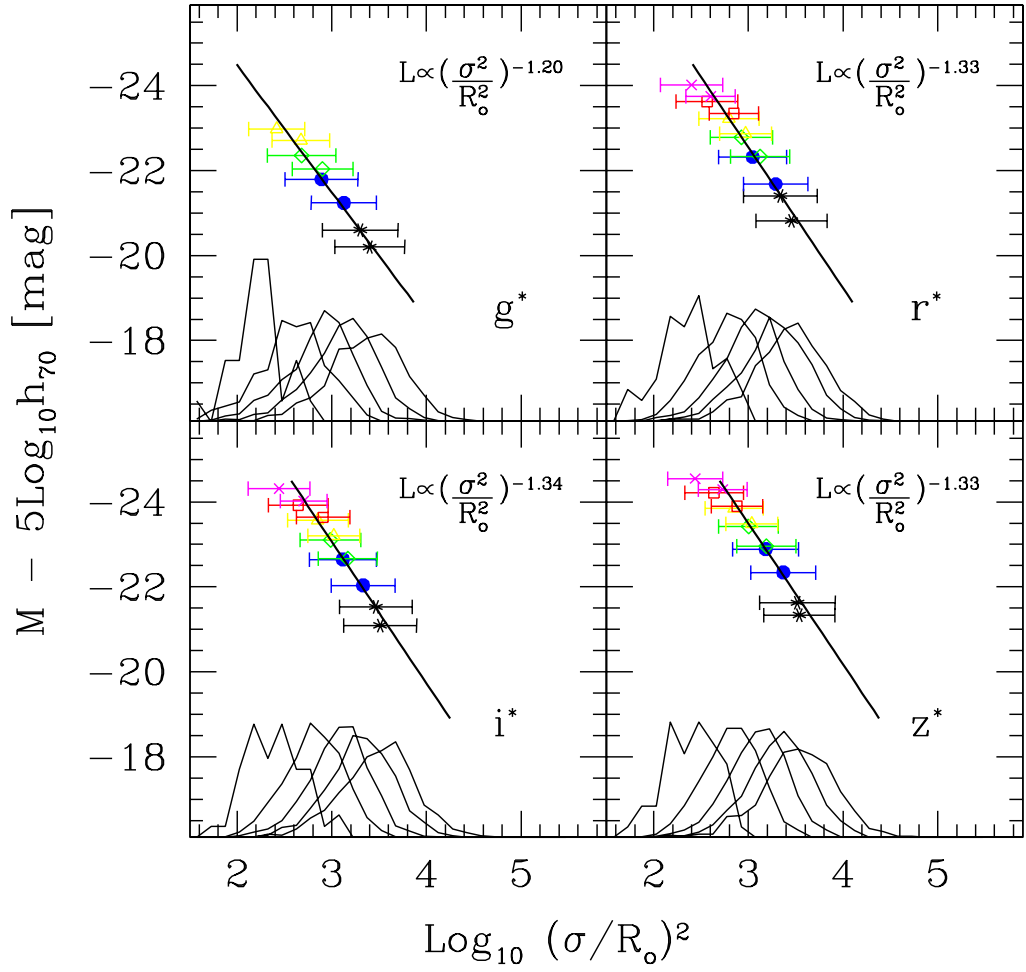}
\vspace{-1.cm}
\caption{Same as previous figure, but for the relation between 
luminosity $L$ and the combination $(\sigma/R_o)^2$, which 
is supposed to be a measure of density.}
\label{lden}
\end{figure}

\begin{figure}
\centering
\epsfxsize=\hsize\epsffile{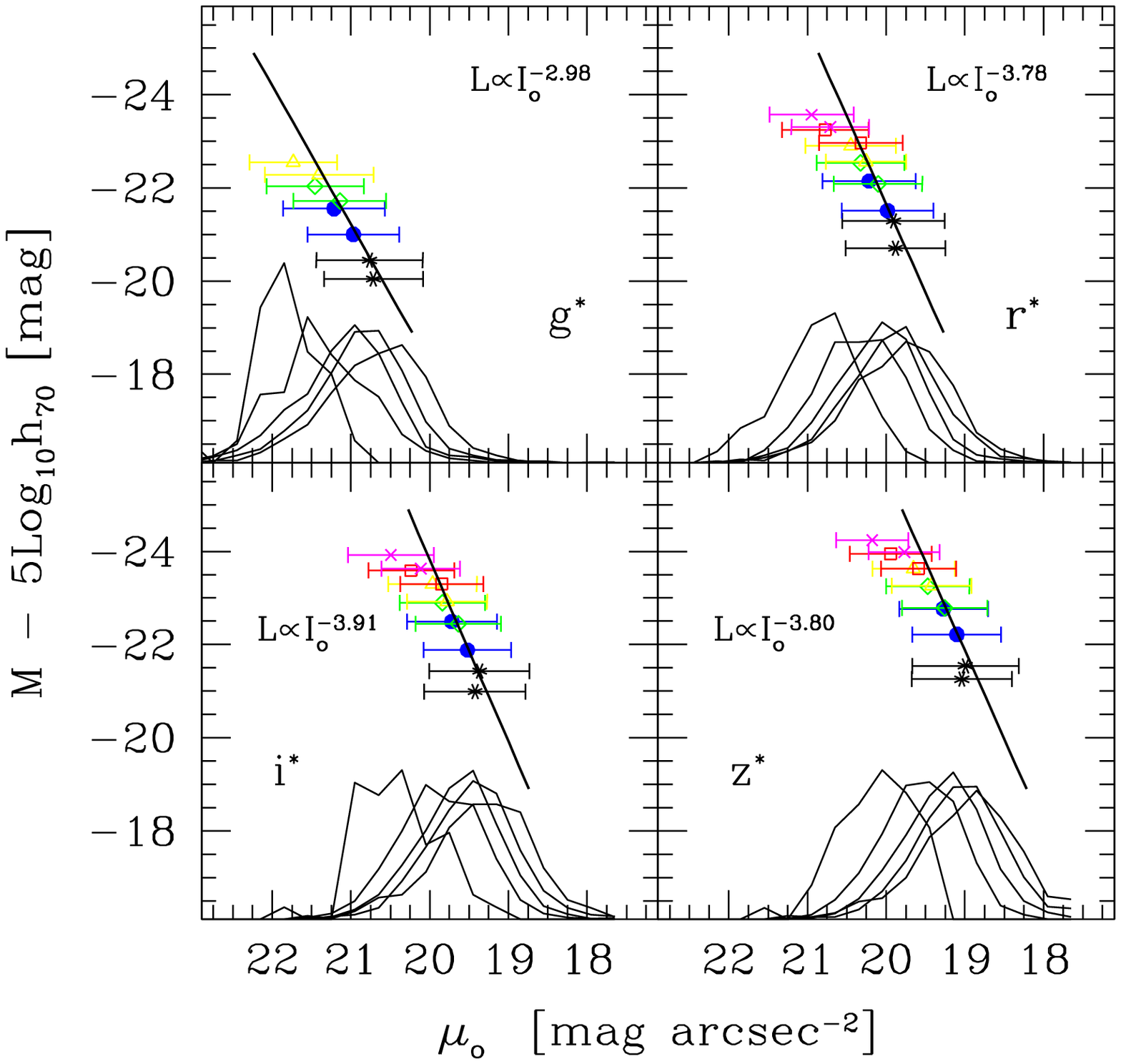}
\vspace{-1.cm}
\caption{Relation between luminosity $L$ and surface brightness $\mu_o$ 
in different volume limited catalogs (higher redshift catalogs contribute 
points to the upper-left corners of each plot).  
Passive evolution of luminosities would shift points upwards and to 
the right of the zero-redshift relation, but, the slope of the relation 
should remain unchanged.  This shift has been subtracted.}
\label{lmu}
\end{figure}

There is an interesting correlation between the residuals of the 
Faber--Jackson and $R_o-L$ relations.  At fixed luminosity, galaxies 
which are larger than the mean $\langle R_o|L\rangle$ tend to have smaller 
velocity dispersions.  
This is shown in Figure~\ref{resid}, which plots the residuals from 
the $\sigma-L$ relation versus the residuals from the $R_o-L$ 
relation.  The short dashed lines show the forward and inverse fits to 
this scatter plot.  The long-dashed line in between the other two shows 
$\Delta_{R|M}/\sigma_{R|M} = -\Delta_{V|M}/\sigma_{V|M}$, 
where $\Delta_{X|M}$ denotes the residual from the mean relation at 
fixed $M$, and $\sigma_{X|M}$ denotes the rms of this residual.  
The anti-correlation is approximately the same for all $L$.  

This suggests that a plot of $L$ versus some combination of $R_o$ and 
$\sigma$ should have considerably less scatter than either of the two 
individual relations.  To illustrate, Figure~\ref{lmass} shows the 
distribution of the combination $R_o\sigma^2$ at fixed $L$.  The scatter 
in $L$ is significantly reduced, making the mean trend of increasing 
$R_o\sigma^2$ with increasing $L$ quite clean.  (The combination of 
observables for which the scatter is minimized is discussed in 
Section~\ref{prjct}.)  This particular combination defines an effective 
mass:  $M_o \equiv 2R_o\sigma^2/G$.  In slightly more convenient units, 
this mass is  
\begin{equation}
\left({M_o\over 10^{10}h^{-1}M_\odot}\right) = 
 0.465\,\left({R_o\over h^{-1}{\rm kpc}}\right)
 \left({\sigma\over 100\,{\rm km~s}^{-1}}\right)^2 .
 \label{meff}
\end{equation}
(Because many of our galaxies are not spherical, some of their support 
must come from rotation, and so ignoring rotation as we are doing is 
likely to mis-estimate the true mass.  
See Bender, Burstein \& Faber 1992 for one way to account for this.  
This quantity will also mis-estimate the mass if some of the support 
comes from anisotropic velocity dispersions.)  

Fiducial values of the effective mass-to-light ratio can be obtained by 
inserting the maximum likelihood values from Table~\ref{MLcov} into this 
relation.  This yields $M_o\approx 10^{10.56}h_{70}^{-1}M_\odot$ (we used 
the parameters for the $r^*$ band, for which $R_*+2V_* \approx 4.89$).  
The corresponding total absolute magnitude is 
$M_*-5\log_{10}h_{70} \approx -21.15$.  
The luminosity of the sun in $r^*$ is $4.62$~mags, so 
$L_*\approx 10^{10.31}h_{70}^{-2}L_\odot$.  
The luminosity within the effective radius is half this value, so that 
the effective mass-to-light ratio within the effective radius of an 
$L_*$ object is $2h_{70}\times 10^{10.56-10.31}\approx 3.57h_{70}$ times 
that of the sun.  
Figure~\ref{lmass} shows that the effective mass-to-light ratio 
depends on luminosity:  
\begin{equation}
 \Big\langle {M_o\over L}\Big|L\Big\rangle 
  = 3.57\,h_{70}\ \left(L\over L_*\right)^{0.15}\ {M_\odot\over L_\odot} 
  \qquad{\rm in}\ r^*.  
\end{equation}
At larger radii, the luminosity can double at most, whereas, if the 
galaxy is embedded in a dark matter halo, the mass at large radii may 
continue to increase.  For this reason one might expect the mass-to-light 
ratios to be significantly larger at larger radii.  

Since the ratio $(2R_o\sigma_o^2/G)/(L/2)$ above is the mass-to-light 
ratio at the radius which encloses half the light, it is tempting to 
associate it with the mass-to-light ratio at the half mass radius.  
Because both the numerator and the denominator are projected quantities, 
this is incorrect.  For example, if the mass-to-light ratio is 
independent of distance from the galaxy center, then the three 
dimensional half-mass radius is about 30\% larger than the projected 
half-light radius (e.g. Hernquist 1990).  If the velocity dispersion 
does not change substantially over the range in radii which contribute 
light, then a fairer estimate of the mass-to-light ratio within the 
half-mass radius would be about 30\% larger than the value given above.  

We can define an effective density by setting 
 $3M_o/4\pi R_o^3 = (2R_o\sigma^2/G)/(4\pi R_o^3/3) 
                  = \Delta_o\,\rho_{\rm crit}$, 
with $\rho_{\rm crit} \equiv 3H^2/8\pi G$, then 
\begin{equation}
\Delta_o = 4\,\left(\sigma\over HR_o\right)^2 = 
 4\times 10^6\times 0.7^2\left({\sigma\over 100\,{\rm km~s}^{-1}}
       {h_{70}^{-1}\ {\rm kpc}\over R_o}\right)^2.
 \label{deff}
\end{equation}
Figure~\ref{lden} shows that this effective density decreases with 
increasing luminosity, although the scatter in densities at fixed 
luminosity is quite large ($\sim 0.32$~dex).  Inserting mean values 
for $\sigma$ and $R_o$ yields 
\begin{equation}
 \Delta_o = 5.16\times 10^5 \,\left({\sigma/\sigma_*\over R_o/R_*}\right)^2
     = 5.16\times 10^5 \,\left({L\over L_*}\right)^{-3/4}
     \qquad {\rm in}\ r^*.  
\end{equation}
Such a trend is qualitatively similar to that seen in numerical 
simulations of dissipationless gravitational clustering:  the central 
densities of virialized halos in such simulations are smaller in the 
more massive halos (Navarro, Frenk \& White 1997).  

Figure~\ref{lmu} shows a final relation at fixed luminosity:  the 
surface-brightness$-L$ relation.  In such a plot, luminosity evolution 
moves objects upwards and to the right (larger luminosities and 
surface brightnesses at high redshift), so that the higher redshift 
population should be obviously displaced from the zero-redshift 
relation.  The plot shows the distribution of $\mu_o$ and $M$ after 
subtracting the maximum likelihood estimate of the evolution from both 
quantities.  The solid line shows the maximum likelihood value of 
the slope of this relation.  This differs slightly from the 
$I\propto L^{-0.45}$ scaling Sandage \& Perelmuter (1990) find for 
giant galaxies with $M_B<-20$, although the scatter around the 
mean relation of $\sim 0.58$ mags is similar.  This relation is 
considerably broader than any of the others we have studied so far, 
which may account for some of the difference.  
However, a careful inspection of the figure suggests that the 
relation is becoming shallower at high redshift; whether or not this 
is a signature of differential evolution in the luminosities is the 
subject of work in progress.

\section{A parametric maximum-likelihood analysis}\label{prjct}
\label{ML3d}
Section~\ref{lf} showed that, after accounting for the fact that the 
SDSS sample is magnitude-limited, the distribution of $M=-2.5\log_{10}L$ 
is quite well described by a Gaussian.  In principle, by extending the 
Efstathiou, Ellis \& Peterson (1988) method (along the lines described 
by Sodr{\'e} \& Lahav 1993) we could derive non-parametric 
maximum-likelihood estimates of the three-dimensional distribution 
of $L$, $R_o$ and $\sigma$.  The virtue of this approach is that it 
accounts for the fact that the observed sample is magnitude-limited, 
that there is also a cut at small velocity dispersions, and that there 
are correlated measurement errors associated with the luminosities, 
sizes and velocity dispersions.  Once the shape of the three-dimensional 
distribution has been estimated, it is straightforward to obtain 
estimates of the various correlations with luminosity we studied in 
the previous section.  This is the subject of Section~\ref{fixedl}.  
However, the real benefit of the maximum likelihood analysis is that 
it also yields estimates of correlations between observables which do 
not include luminosity---some examples of these are shown in 
Section~\ref{others}.  

We chose not to make a non-parametric estimate of the joint distribution 
because just ten bins in each of $L$, $R_o$ and $\sigma$ yields $10^3$ 
free parameters to be determined from $10^4$ galaxies.  Moreover,  
Section~\ref{lx} showed that, in each of the SDSS wavebands, the 
distributions of $\log_{10}R_o$ and $\log_{10}\sigma$ at fixed absolute 
magnitude are quite well described by Gaussian forms.  Therefore, the 
joint distribution of early-type galaxy luminosities, sizes, and 
velocity dispersions should be well described by a tri-variate Gaussian 
distribution in the variables $M=-2.5\log_{10}L$, $R=\log_{10}R_o$ and 
$V=\log_{10}\sigma$.  Saglia et al. (2001) describe a maximum likelihood 
analysis of early-type galaxy correlations in which they {\it assume} 
that a tri-variate Gaussian is a reasonable description of their data:  
we have the luxury of knowing that this is indeed a reasonable description 
of our dataset.  Thus, we have a simple parametrization of the joint 
distribution for which, in each waveband, nine numbers suffice to 
describe the statistical properties of our sample:  
three mean values, $M_*$, $R_*$ and $V_*$, 
three dispersions, $\sigma^2_M$, $\sigma^2_R$ and $\sigma^2_V$, 
and three pairwise correlations, $\sigma_R\sigma_M\,\rho_{RM}$, 
$\sigma_V\sigma_M\,\rho_{VM}$, and $\sigma_R\sigma_V\,\rho_{RV}$.  

In addition, we will also allow for the possibility that the 
luminosities are evolving---a tenth parameter to be estimated from 
the sample.  The maximum likelihood technique allows us to estimate 
these ten numbers as follows.  We define the likelihood function  
\begin{eqnarray}
{\cal L} &=& \prod_i {\phi({\cal X}_i,{\cal C},{\cal E}_i)\over S(z_i)},
 \qquad{\rm where}\nonumber \\
{\cal X} &=& (M-M_*+Qz,R-R_*,V-V_*) , \nonumber\\
{\cal E} &=& \left( \begin{array}{ccc} 
                      \epsilon^2_{MM} & \epsilon^2_{RM} & \epsilon^2_{VM} \\ 
                      \epsilon^2_{RM} & \epsilon^2_{RR} & \epsilon^2_{RV} \\
                      \epsilon^2_{VM} & \epsilon^2_{RV} & \epsilon^2_{VV} \\
                    \end{array}\right),\nonumber\\
{\cal C} &=& \left( \begin{array}{ccc} 
     \sigma^2_M & \sigma_R\sigma_M\,\rho_{RM} & \sigma_V\sigma_M\,\rho_{VM}\\
     \sigma_R\sigma_M\,\rho_{RM} & \sigma^2_R & \sigma_R\sigma_V\,\rho_{RV}\\
     \sigma_V\sigma_M\,\rho_{VM} & \sigma_R\sigma_V\,\rho_{RV} & \sigma^2_V\\
           \end{array}\right)\qquad{\rm and} \nonumber\\
\phi({\cal X},{\cal C},{\cal E}) &=& 
{\phi_*\over (2\pi)^{3/2}\,|{\cal C}+{\cal E}|^{-1/2}}\, 
\exp\left(-{1\over 2}\,{\cal X}^{T}\,[{\cal C}+{\cal E}]^{-1} {\cal X}\right).
\end{eqnarray}
Similarly to when we discussed the luminosity function, $S(z_i)$ is 
defined by integrating over the range of absolute magnitudes, velocities 
and sizes at $z_i$ which make it into the catalog.  Here ${\cal X}$ is 
the vector of the observables, and ${\cal E}$ describes the errors in 
the measurements.  

Appendix~D of Paper~I describes how the elements of the error matrix 
${\cal E}$ were obtained.  Briefly, the error in the absolute 
magnitude assumes that there are no errors in the redshift or the 
K-correction, so all the error comes from the error on the apparent 
magnitude $m_{\rm dev}$; 
the error on the circularly averaged radius $R_o$ is given by adding 
the error on the angular length of the longer axis $r_{\rm dev}$ to 
those which come from the error on the axis ratio $b/a$.  
We assume that the errors in $b/a$ are neither correlated with those 
in $\log_{10}r_{\rm dev}$ nor with those in the absolute magnitude.  
However, because both $m_{\rm dev}$ and $r_{\rm dev}$ come from the 
same fitting procedure, the errors in $M$ and $R_o$ are correlated.  
Finally, we assume that errors in magnitudes are not correlated 
with those in velocity dispersion, so $\epsilon^2_{VM}$ 
is set to zero, and that errors in size and velocity dispersion are 
only weakly correlated because of the aperture correction we apply.  

\begin{table}[t]
\centering
\caption[]{Maximum-likelihood estimates, in the four SDSS bands, of the 
joint distribution of luminosities, sizes and velocity dispersions.  
The mean values of the variables at redshift $z$, 
$M_*-Qz$, $R_*$, $V_*$, and the elements of the covariance matrix 
${\cal C}$ defined by the various pairwise correlations between the 
variables are shown.  These coefficients are also used 
in computing the matrix ${\cal F}$ in Paper~III.\\}
\begin{tabular}{cccccccccccc}
\tableline 
Band & $N_{\rm gals}$ & $M_*$ & $\sigma_M$ & $R_*$ & $\sigma_R$ &
$V_*$ & $\sigma_V$ & $\rho_{RM}$ & $\rho_{VM}$ & $\rho_{RV}$ & Q\\
%     & mag & mag & dex & dex & dex & dex & & & \\
\hline\\
$g^*$ & 5825 & $-20.43$ & 0.844 & 0.520 & 0.254 & 2.197 & 0.113 & $-0.886$ &
$-0.750$ & 0.536 & 1.15 \\
$r^*$ & 8228 & $-21.15$ & 0.841 & 0.490 & 0.241 & 2.200 & 0.111 & $-0.882$ &
$-0.774$ & 0.543 & 0.85 \\
$i^*$ & 8022 & $-21.49$ & 0.851 & 0.465 & 0.241 & 2.201 & 0.110 & $-0.886$ &
$-0.781$ & 0.542 & 0.75 \\
$z^*$ & 7914 & $-21.83$ & 0.845 & 0.450 & 0.241 & 2.200 & 0.110 & $-0.885$ &
$-0.782$ & 0.543 & 0.60 \\
\tableline
\end{tabular}
\label{MLcov}
\end{table}

The covariance matrix ${\cal C}$ contains six of the ten free parameters 
we are seeking.  It is these parameters, along with the three mean values, 
$M_*$, $R_*$ and $V_*$, and the evolution parameter $Q$ which are varied 
until the likelihood is maximized.  The maximum-likelihood estimates of 
these parameters in each band are given in Table~\ref{MLcov}.  
Notice that although the luminosity and size distributions differ from 
band to band, the velocity distributions do not.  This is reassuring, 
because the intrinsic distribution of velocity dispersions, estimated 
from the spectra, should not depend on the band in which the photometric 
measurements were made.  
As an additional test, we also computed maximum-likelihood estimates 
of the $2\times 2$ covariance matrices of the bivariate Gaussians for 
the pairs $(M,R)$ and $(M,V)$.  These estimates of, e.g., $\rho_{RM}$ 
and $\rho_{VM}$ were similar to those in Table~\ref{MLcov}.  

The remainder of this paper uses ${\cal C}$ to estimate various 
pairwise correlations.  
In Paper~III, we transform the covariance matrix ${\cal C}$ into one 
which describes the Fundamental Plane variables of size, surface 
brightness and velocity dispersion.  

\subsection{The intrinsic distributions of sizes and velocity dispersions}
Before we present maximum likelihood estimates of various correlations, 
it is worth remarking that because the trivariate Gaussian is a good 
description of the data, our results indicate that, in addition to 
the intrinsic distribution of absolute magnitudes, the intrinsic 
distributions of (the logarithms of) early-type galaxy sizes and 
velocity dispersions are also well fit by Gaussian forms.  
The means and dispersions of these Gaussians are given by 
$(R_*,\sigma^2_R)$ and $(V_*,\sigma^2_V)$ in Table~\ref{MLcov}.  
Note that the width of the distribution of $\log_{10}\sigma$ is 
about half that of $\log_{10}R_o$.  This is consistent with earlier 
work (e.g., it is one of the motivations for the $\kappa$-space 
parametrization of Bender, Burstein \& Faber 1992).  

\subsection{Correlations with luminosity}\label{fixedl}
As we describe below, appropriate combinations of the coefficients in 
Table~\ref{MLcov} provide maximum likelihood estimates of various 
linear regressions between pairs of observables which are often 
studied; these are summarized in Table~\ref{lx4cov}.  Plots comparing 
some of these linear regressions with the maximum likelihood estimates 
are shown in Section~\ref{lx}.  

In the Gaussian model, the mean of $\log_{10}\sigma$ at fixed $M$ is 
\begin{equation}
\Bigl\langle V-V_*|M-M_*\Bigr\rangle = 
 {(M-M_*)\over\sigma_M}\ \sigma_V\,\rho_{VM} 
 \equiv {(M-M*)\over -2.5\,S_{VL}},
\label{meanvm}
\end{equation}
where the second equality defines $S_{VL}$, for ease of comparison 
with equation~(\ref{xlregress}).  The dispersion around this mean is 
\begin{equation}
 \sigma^2_{V|M}\equiv \sigma^2_{V}(1 - \rho^2_{VM}).  
\label{varvm}
\end{equation}
Inserting the values in Table~\ref{MLcov} into these expressions for 
$S_{VL}$ and $\sigma^2_{V|M}$ provides the maximum likelihood 
estimate of the slope and thickness of this relation.  These are 
shown in the second column of Table~\ref{lx4cov}, and the fit 
itself is shown in Figure~\ref{ls}.  
The errors we quote on the slopes of this, and the other relations 
in the Table, were obtained using subsamples as described in 
Appendix~\ref{compcat}.  Note that the errors we find in this way are 
comparable to those sometimes quoted in the literature, even though 
each of the subsamples we selected is an order of magnitude larger than 
any sample available in the literature.  

\begin{table}[t]
\centering
\caption[]{Maximum-likelihood estimates of the slopes 
$S_{VL}$, $S_{RL}$, $S_{ML}$, $S_{DL}$, and $S_{IL}$, 
of the relations between luminosity and the mean velocity dispersion, 
effective radius, effective mass, effective density, and effective 
surface brightness at fixed luminosity, as a function of luminosity.  
The slope of the relation between surface brightness and the mean size 
at fixed surface brightness is $S_k$.\\}
\begin{tabular}{ccccccc}
\tableline
Band  & $S_{VL}$ & $S_{RL}$ & $S_{ML}$ & $S_{DL}$ & $S_{IL}$ & $S_k$ \\
\tableline\\
$g^*$ & 4.00$\pm 0.25$ & 1.50$\pm 0.06$ & 0.86$\pm 0.02$ & $-1.20\pm 0.08$ & $-2.98\pm 0.16$ & $-0.73\pm 0.02$ \\
$r^*$ & 3.91$\pm 0.20$ & 1.58$\pm 0.06$ & 0.87$\pm 0.02$ & $-1.33\pm 0.07$ & $-3.78\pm 0.17$ & $-0.75\pm 0.02$ \\
$i^*$ & 3.95$\pm 0.15$ & 1.59$\pm 0.06$ & 0.88$\pm 0.02$ & $-1.34\pm 0.08$ & $-3.91\pm 0.18$ & $-0.76\pm 0.02$ \\
$z^*$ & 3.92$\pm 0.15$ & 1.58$\pm 0.06$ & 0.88$\pm 0.02$ & $-1.33\pm 0.07$ & $-3.80\pm 0.17$ & $-0.76\pm 0.01$ \\
\tableline \\
\end{tabular}
\label{lx4cov}
\end{table}

% Notice that $\sigma_{V|M}$ is not negligible compared to $\sigma_V$.  
% This has an important consequence.  The distribution of velocity 
% dispersions is sometimes (e.g., when spectroscopic data are not available) 
% approximated by substituting the mean Faber--Jackson relation in the 
% expression for the luminosity function.  This simple change of variables 
% is only accurate if the scatter in the Faber--Jackson relation is 
% negligible---for the galaxies in our dataset, this is not the case
% (see Figure~\ref{ls}).  
% Because the simple change of variables underestimates the width of the 
% velocity dispersion distribution, it greatly underestimates the number 
% density of galaxies which have large velocity dispersions.  

The mean size at fixed absolute luminosity $M$, and the dispersion around 
this mean, are obtained by replacing all $V$'s with $R$'s in 
equation~(\ref{meanvm}).  
The third column in Table~\ref{lx4cov} gives the maximum likelihood 
value of the slope $S_{RL}$, of the size-at-fixed-luminosity relation 
in the four bands.  This fit is shown in Figure~\ref{lre}.  
% As was the case with the velocities, approximating the distribution of 
% sizes by using the size-luminosity relation to change variables in the 
% luminosity function is not particularly accurate, although, because 
% $\rho_{RM}$ is larger than $\rho_{VM}$, the approximation is slightly 
% better for the sizes than for the velocities.  

Similarly, one can show that the slopes of the mean $L$-mass and 
$L$-density relations shown in Figures~\ref{lmass} and~\ref{lden} 
are $S_{ML} = (2/S_{VL} + 1/S_{RL})^{-1}$ and 
$S_{DL} = 1/(2/S_{VL} - 2/S_{RL})^{-1}$.  These are the fourth and 
fifth columns of Table~\ref{lx4cov}.  The dispersions around these 
mean mass-$L$ and density-$L$ relations can be written in terms of 
the elements of ${\cal C}$, though we have not included the expressions 
here.  Even though these relations are made from linear combinations of 
$R$ and $V$, they may be tighter than either the $L-\sigma$ or $L-R_o$ 
relations because the correlation coefficients $\rho_{RM}$, $\rho_{VM}$ 
and $\rho_{RV}$ are different from zero.  

The surface brightnesses of the galaxies in our sample are defined 
by $(\mu_o-\mu_*) \equiv (M-M_*) + 5 (R-R_*)$, so the dispersion in 
$\mu$ is 
$\sigma_{\mu}^2 = 
 \sigma_M^2 + 10\sigma_M\sigma_R\rho_{RM} + 25\sigma_R^2$.  
The mean surface brightness at fixed luminosity is obtained by 
replacing all $V$s with $\mu$s in the equations~(\ref{meanvm}) 
and~(\ref{varvm}) above.  
This means that we need $\rho_{\mu M}$, which we can write in 
terms of $\sigma_M$, $\sigma_R$ and $\rho_{RM}$.  
The sixth column in Table~\ref{lx4cov} gives the slope of the surface 
brightness $I$ at fixed luminosity relation, 
$\langle I_o|L\rangle\propto L^{1/S_{IL}}$, in the four bands.  
These fits are shown in Figure~\ref{lmu}.  

\subsection{Inverse relations and other correlations}\label{others}
So far, we have shown that the maximum likelihood analysis provides 
estimates of correlations which are in good agreement with quantities 
which can also be estimated by a more straightforward regression 
technique.  However, with the coefficients of the correlation matrix 
${\cal C}$ in hand, it is straightforward to obtain estimates of 
correlations which, because of selection effects, cannot be reliably 
estimated using simple regressions.  For example, the mean luminosity 
given the velocity dispersion is 
\begin{equation}
\Bigl\langle M-M_*|V-V_*\Bigr\rangle = 
 {(V-V_*)\over\sigma_V}\ \sigma_M\,\rho_{VM} 
\end{equation}
with dispersion $\sigma_M^2(1-\rho_{VM}^2)$ 
(compare equations~\ref{meanvm} and~\ref{varvm}).  
Inserting the coefficients in Table~\ref{MLcov} yields 
$\langle L|\sigma\rangle\propto \sigma^{2.34}$.  Similarly, one can 
show that $\langle L|R_o\rangle\propto R_o^{1.23}$ and 
$\langle M_o/L|M_o\rangle \propto M_o^{0.22}$ in $r^*$.  

We can also study correlations which do not involve luminosity.  
The best studied of these is the Kormendy (1977) relation:  
the surface brightnesses of early-type galaxies decrease with 
increasing effective radius.  
The mean size at fixed surface brightness in our sample is 
\begin{equation}
\Bigl\langle R-R_*\Big|\mu-\mu_*\Bigr\rangle = 
{(\mu-\mu_*)\over \sigma_{\mu}}\ \sigma_R\,\rho_{\mu R} \equiv 
    -0.4\,S_k\,(\mu-\mu_*).
\end{equation}
where $\rho_{\mu R}$ can be written in terms of $\sigma_M$, $\sigma_R$ 
and $\rho_{RM}$, and the final equality defines $S_k$.  
The seventh column in Table~\ref{lx4cov} gives the slope of this 
relation in the four bands.  For comparison, Kormendy (1977) found 
that $\log_{10}I_o \propto 1.29 \log_{10} R_o$ in the B-band, and 
Pahre et al. (1998) find $R_o\propto I_o^{-0.61}$ in the K-band.  

\begin{figure}
\centering
\epsfxsize=\hsize\epsffile{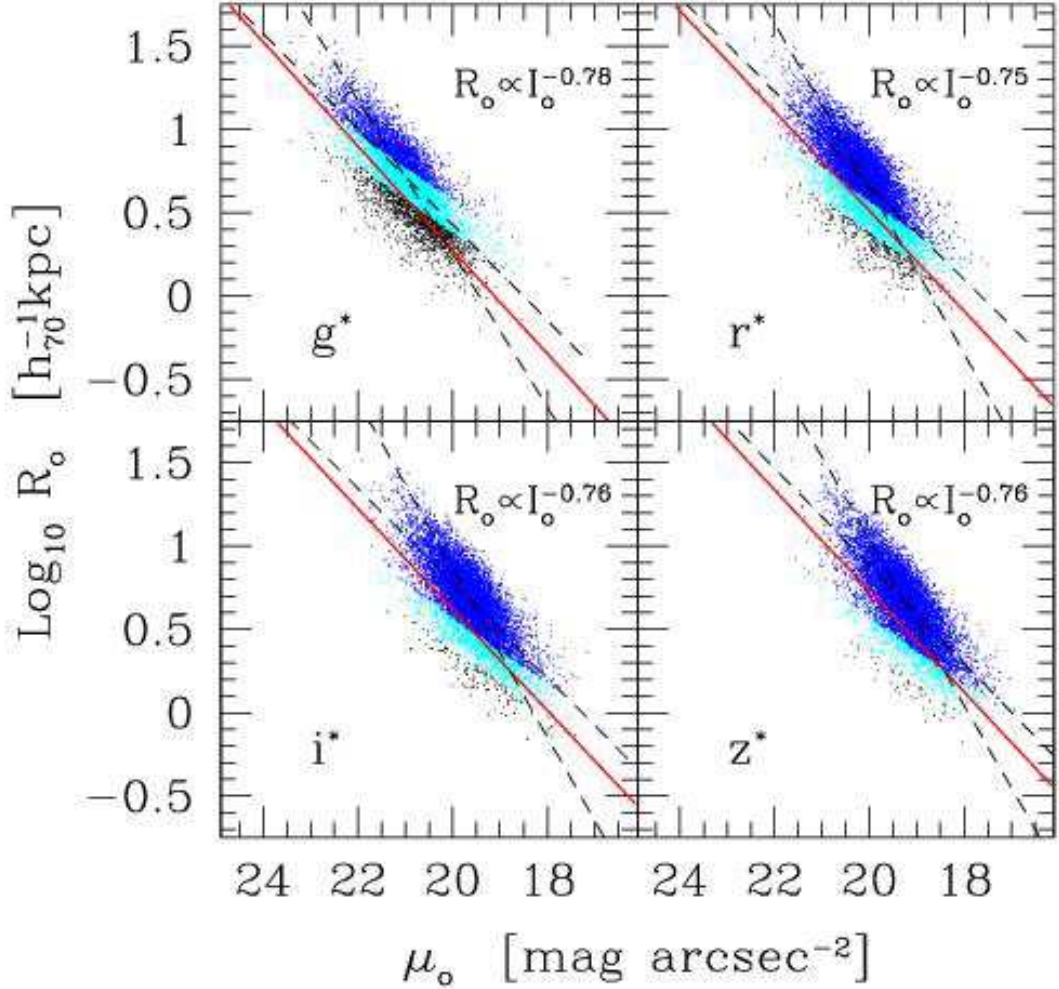}
\vspace{0.cm}
\caption{Relation between effective radius and surface brightness.  
Short dashed lines show forward and inverse fits to this relation.  
The zero-points of these fits are strongly affected by the magnitude 
limit of our sample.  To illustrate, solid line shows the 
maximum-likelihood estimate of the relation in the simulated 
complete catalog from which the magnitude limited catalog, shown 
by the dotted line, was drawn.  }
\label{krmndy}
\end{figure}

For the reasons described in Section~\ref{lx}, when presented with a 
magnitude limited catalog, correlations at fixed luminosity are useful 
because they are unbiased by the selection.  When luminosity is not one 
of the variables then forward and inverse correlations may be equally 
interesting, and equally biased.  For example, in the Kormendy (1977) 
relation, $\langle R-R_*\Big|\mu-\mu_*\rangle$ may be just as interesting 
as $\langle\mu-\mu_*|R-R_*\rangle$.  The slopes of the two relations 
are, of course, simply related to each other.  In fact, it may be 
preferable to study the relations which are defined by the principle 
axes of the ellipse in $(R,\mu)$ space which the galaxies populate.  
The directions of these axes are obtained by computing the eigenvalues 
and vectors of the covariance matrix associated with the sizes and 
surface brightnesses.  To illustrate, the eigenvalues of the $2\times 2$ 
covariance matrix associated with the Kormendy relation are 
\begin{displaymath}
\sigma^2_{\pm} = \Bigl( \sigma_R^2 + \sigma_\mu^2 \pm\sqrt{D_{\mu R}} \Bigr)
\Big/2,
\end{displaymath}
where we have set 
$D_{\mu R} = 
 (\sigma_R^2 - \sigma_\mu^2)^2 + (2\sigma_R\sigma_\mu\rho_{\mu R})^2$.
The $+/-$ eigenvalues give the dispersions along and perpendicular to 
the major axis of the ellipse.  
The long axis of the ellipse describes the mean relation, 
$(R-R_*) = S_K\,(\mu_o - \mu_*)$, where 
\begin{displaymath}
S_K = {\sigma_R^2 - \sigma_\mu^2 + \sqrt{D_{\mu R}}\over 
               2\sigma_R\sigma_\mu\,\rho_{\mu R}}.
\end{displaymath}
With obvious changes of variables, analogous expressions can be derived 
for all the correlations presented earlier, although we do not show 
them here.  

The Kormendy relation in our sample is shown in Figure~\ref{krmndy}.  
The dashed lines show forward and inverse fits to the data:
i.e., the mean size at fixed surface brightness, and the mean 
surface brightness at fixed size.  The parameters of the fits are 
affected by the magnitude limit of the catalog.  
To estimate the effect of the magnitude limit cut on this relation,
we compute the direct and inverse fits to the Kormendy relation in 
the simulated complete and magnitude-limited samples we describe in 
Appendix~\ref{simul}.  The dotted line in Figure~\ref{krmndy} shows 
the direct fit to the magnitude limited simulations (it can hardly 
be distinguished from the fit to the data).  

In comparison, the maximum-likelihood estimate of the true direct 
relation provides a very good description of the relation in the 
complete simulations in which there is no magnitude limit: it is 
shown as the solid line.  Notice that the dashed and dotted lines 
have approximately the same slope as the solid line:  the magnitude 
limit hardly affects the slope, although it changes the zero-point 
dramatically.  At fixed surface brightness, the typical $R_o$ is 
significantly larger in the magnitude limited sample than in the 
complete sample.  This happens because lines of constant luminosity 
run downwards and to the right with slope $-1/5$, so that changes in 
luminosity act approximately perpendicular to the relation.  

This shows that although linear regression fits to the data provide 
good estimates of the true slope of the Kormendy relation, they 
provide bad estimates of the true zero-point.  In comparison, the 
maximum-likelihood technique, which accounts for the selection on 
apparent magnitudes, is able to estimate the slope and the zero-point 
correctly.  

\begin{figure}
\centering
\epsfxsize=\hsize\epsffile{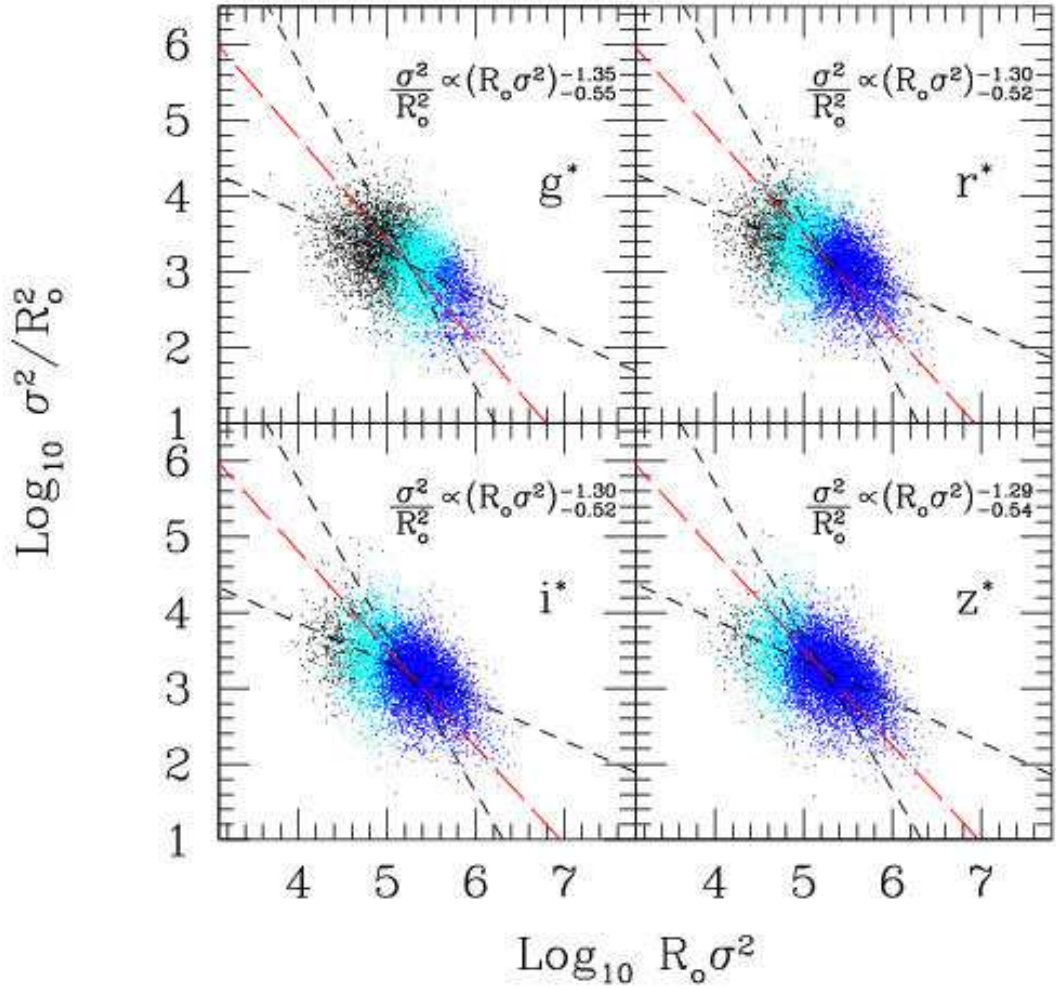}
\vspace{0.cm}
\caption{Relation between effective density and effectve mass.  
Short dashed lines show forward and inverse fits to this relation.  
The more massive galaxies are less dense.  Text in top right of 
each panel shows the slopes of the dashed lines. }
\label{mden}
\end{figure}

Another interesting correlation is that between the effective mass and 
density defined in equations~(\ref{meff}) and~(\ref{deff}).  A little 
algebra shows that $\langle \Delta_o|M_o\rangle\propto M_o^{-0.52}$.  
Figure~\ref{mden} shows forward and inverse fits to this relation.  
The characteristic density $\Delta_c$ of halos seen in numerical 
simulations of hierarchical clustering scales with halo mass:  
 $\langle \Delta_c|m\rangle \sim 9^3\,(m/m_*)^{-0.4}$ 
(Bullock et al. 2001), which is qualitatively similar to the scaling 
of effectve density with effective mass in our sample.  The scatter 
in characteristic densities at fixed halo mass, $\sim 0.33$~dex, is 
also rather similar to the scatter in effective densities at fixed 
mass.  These coincidences may provide important clues to how early-type 
galaxies formed.  

\begin{table}[t]
\centering
\caption[]{Coefficients $\alpha$ and $\beta$ which define the projection 
of minimum scatter, $\sigma_{MRV}$, in the space defined by absolute 
magnitude, and the logarithms of the size and velocity dispersion.\\}
\begin{tabular}{cccc}
\tableline
Band & $\alpha$ & $\beta$ & $\sigma_{MRV}$ \\
\tableline\\
$g^*$ & 0.76 & 1.94 & 0.063 \\
$r^*$ & 0.79 & 1.93 & 0.058 \\
$i^*$ & 0.82 & 1.89 & 0.054 \\
$z^*$ & 0.81 & 1.90 & 0.054 \\
\tableline \\
\end{tabular}
\label{FPlvr}
\end{table}

In contrast to the Faber--Jackson, radius--luminosity, Kormendy, 
and mass--density relations, the relations between luminosity and mass 
and luminosity and density involve three variables.  Is there some 
combination of these variables which provides the least scatter?  
The eigenvectors and eigenvalues of the $3\times 3$ matrix ${\cal C}$ 
give the directions of the principle axes of the ellipsoid in $(M,R,V)$ 
space which the early-type galaxies populate.  
One of the eigenvalues of ${\cal C}$ is considerably smaller than the 
others, suggesting that the galaxies populate a two-dimensional plane 
in $(M,R,V)$ space.  The eigenvectors show that the plane is viewed 
edge-on in the projection 
\begin{equation}
-0.4(M-M_*) = \alpha\,(R-R_*) + \beta\,(V-V_*),
\end{equation}
where $M_*$, $R_*$ and $V_*$ were given in Table~\ref{MLcov}, and the 
coefficients $\alpha$ and $\beta$, and the thickness of the plane in 
this projection, $\sigma_{MRV}$, are given in Table~\ref{FPlvr}.  
Section~\ref{lx} shows that a scatter plot of luminosity versus mass 
is considerably tighter than plots of $M$ versus $\log_{10}R_o$ or 
$\log_{10}\sigma$.  The eigenvectors of ${\cal C}$ show that this is 
because the $M$ versus $R + 2V$ projection is actually quite close to 
the edge-on projection.  
It is interesting that this plane is only about 10\% thicker than 
the Fundamental Plane relation between $R_o$, $I_o$ and $\sigma$ 
which is the subject of Paper~III.

\section{Discussion and conclusions}\label{discuss}
We have studied the properties of $\sim 9000$ early-type galaxies 
over the redshift range $0\le z\le 0.3$ using photometric (the 
$g^*$, $r^*$, $i^*$ and $z^*$ bands) and spectroscopic observations.  
The intrinsic distributions of luminosity, velocity dispersion and 
half-light radius of the galaxies in our sample are each well 
described by Gaussians in absolute magnitude, $\log_{10}\sigma$, 
and $\log_{10}R_o$.  

A maximum likelihood analysis of the joint distribution of luminosities, 
sizes and velocity dispersions suggests that the population at higher 
redshifts is slightly brighter than the population nearby, and that 
the change with redshift is faster in the shorter wavebands:  
If $M_*(z) = M_*(0) - Qz$, then $Q=1.15$, $0.85$ and $0.75$ in 
$g^*$, $r^*$ and $i^*$.  
This evolution is sufficiently weak that, relative to their values at 
the median redshift ($z\sim 0.15$) of our sample, the sizes, surface 
brightnesses and velocity dispersions of the early-type galaxy population 
at lower and higher redshifts has evolved little.  
The fact that we see so little evolution in the luminosities argues 
for a relatively high formation redshift:  Bruzual \& Charlot (2003) 
single burst stellar population synthesis models indicate that 
$t_{\rm form}\sim 9$~Gyrs.  This is consistent with the model we use 
to make K-corrections in Paper~I, and is also consistent with the 
formation time estimates based on the Fundamental Plane in Paper~III, 
and galaxy colors and spectral line indices in Paper~IV.    

We find that $\langle\sigma|L\rangle \propto L^{1/4}$ and 
$\langle R_o|L\rangle\propto L^{3/5}$ (see Table~\ref{lx4cov} for 
the exact coefficients, and Figures~\ref{ls} and~\ref{lre} for the 
fits).  Galaxies which are slightly larger than expected (given their 
luminosity) have smaller velocity dispersions than expected 
(Figure~\ref{resid}).  This is expected if galaxies are in virial 
equilibrium.  

A plot of luminosity versus effective mass $M_o = 2R_o\sigma^2/G$  
is substantially tighter than either the $L-\sigma$ or the $L-R_o$ 
relations.  It has a slope which is slightly shallower than unity.  
In particular, on scales of a few kiloparsecs, $L\propto M_o^{0.86}$, 
approximately independent of waveband (Figure~\ref{lmass}).  This 
complements recent SDSS weak-lensing analyses (McKay et al. 2001) 
which suggest that mass is linearly proportional to luminosity in 
these same wavebands, but on scales which are two orders of magnitude 
larger ($\sim 260h^{-1}$kpc).  Together, these two measurements of 
the mass-to-light ratio can be used to provide a constraint on the 
density profiles of dark matter halos.  

A plot of luminosity versus effective density 
$\Delta_o\propto \sigma^2/R_o^2$ shows that 
$\langle\Delta_o|L\rangle\propto L^{-3/4}$ (Figure~\ref{lden}).  
Moreover, a maximum likelihood analysis suggests that the more massive 
galaxies are less dense:  $\langle\Delta_o|M_o\rangle\propto M_o^{-0.52}$
(Figure~\ref{mden}).  
This is qualitatively similar to a trend seen in numerical simulations 
of hierarchical clustering: more massive halos tend to be less centrally 
concentrated (Navarro, Frenk \& White 1997).  
This coincidence may provide an important clue to how early-type 
galaxies formed.  
% The central densities of halos are 
% fixed by when they formed.  Since this likely coincides with the 
% epoch when most of the stars formed, this suggests a formation picture 
% in which dissipation must have been important (to increase the 
% effective densities), but that otherwise a simple linear scaling of halo 
% scale radius to half-light radius is a reasonable approximation.  

The Kormendy relation between size and surface brightness has 
approximately the same slope $\langle R_o|I_o\rangle \propto I_o^{-0.77}$ 
in all four SDSS bands (Figure~\ref{krmndy}).  
Our maximum likelihood analysis, and measurements made in mock catalogs 
which reproduce all the observed scalings of the dataset (a procedure 
for generating such catalogs is described in Appendix~\ref{simul}), 
show that the zero-point of this relation is strongly affected by the 
magnitude limit of the sample (Section~\ref{others}).

\vspace{1cm}

Funding for the creation and distribution of the SDSS Archive has been 
provided by the Alfred P. Sloan Foundation, the Participating Institutions, 
the National Aeronautics and Space Administration, the National Science 
Foundation, the U.S. Department of Energy, the Japanese Monbukagakusho, 
and the Max Planck Society. The SDSS Web site is http://www.sdss.org/.

The SDSS is managed by the Astrophysical Research Consortium (ARC) for
the Participating Institutions. The Participating Institutions are The
University of Chicago, Fermilab, the Institute for Advanced Study, the
Japan Participation Group, The Johns Hopkins University, Los Alamos
National Laboratory, the Max-Planck-Institute for Astronomy (MPIA),
the Max-Planck-Institute for Astrophysics (MPA), New Mexico State
University, University of Pittsburgh, Princeton University, the United
States Naval Observatory, and the University of Washington.

{}

\appendix 

\section{Simulating a complete sample}\label{simul}
This Appendix describes how to use our knowledge of the covariance 
matrix ${\cal C}$ to simulate mock galaxy samples which have the 
same correlated observables as the data.  We use these mock 
samples to estimate the effect of the magnitude limit cut on the  
relations we wanted to measure in the main text.  

The observed parameters $L$, $R_o$ and $\sigma$ of each galaxy in our 
sample are drawn from a distribution, say, $\phi(M,R,V|z)$, where 
$M$ is the absolute magnitude, $R=\log_{10}R_o$ and $V=\log_{10}\sigma$.  
We show in Section~\ref{lx} that $\phi(M,R,V|z) = p(R,V|M,z)\,\phi(M|z)$, 
where $\phi(M|z)$ is the luminosity function at redshift $z$, and the 
distribution of $R$ and $V$ at fixed luminosity is, to a good approximation, 
a bivariate Gaussian.  The maximum likelihood estimates of the parameters 
of the luminosity function and of the bivariate distribution at fixed 
luminosity can be obtained from Table~\ref{MLcov}.

To make the simulations we must assume that, when extrapolated down to 
luminosities which we do not observe, these relations remain accurate.  
Assuming this is the case, we draw $M$ from the Gaussian distribution 
that we found was a good fit to $\phi(M|z)$ (Section~\ref{lf}).  
We then draw $R$ from the Gaussian distribution with mean 
$\langle R|M\rangle$ and dispersion $\sigma_{R|M}^2$.  
Finally, we draw $V$ from a Gaussian distribution with mean and variance 
which accounts for the correlations with both $M$ and $R$.  In practice 
we draw three zero mean unit variance Gaussian random numbers:  
$g_0$, $g_1$, and $g_2$, and then set 
\begin{eqnarray*}
M &=& M_* + \sigma_M\,g_0,\nonumber\\
R &=& R_* + {(M-M_*)\over\sigma_M}\ \sigma_R\,\rho_{RM} +
g_1\,\sigma_R\sqrt{1-\rho_{RM}^2} \qquad {\rm and} \nonumber\\
V &=& V_* + {(M-M_*)\over\sigma_M}\,\xi_{MV} 
          + {(R-R_*)\over\sigma_R}\,\xi_{RV} + g_2\,\sigma_{V|RM},
          \qquad {\rm where} \nonumber\\
\xi_{MV} &=& \sigma_V
{(\rho_{VM} - \rho_{RM}\rho_{RV})\over (1 - \rho_{RM}^2)},\nonumber\\
\xi_{RV} &=& \sigma_V {(\rho_{RV}-\rho_{RM}\rho_{VM})\over (1-\rho_{RM}^2)}, 
\qquad {\rm and} \nonumber\\
\sigma_{V|RM} &=& \sigma_V \sqrt{{1 - \rho_{RM}^2 - \rho_{RV}^2 - \rho_{VM}^2 +
2\rho_{RV}\rho_{VM}\rho_{RM}\over 1 - \rho_{RM}^2}}.
\end{eqnarray*}
Because each simulated galaxy is assigned a luminosity and size, 
its surface brightness is also fixed:  $\mu = M + 5R + $ constant.  

If we generate a catalog in $r^*$, then we can also generate colors 
using the parameters given in Table~1 of Paper~IV.  
Specifically, generate a Gaussian variate $g_3$, and then set 
 $C = C_* + \xi_{CM}(M-M_*)/\sigma_M + \xi_{CV}(V-V_*)/\sigma_V + 
      g_3\,\sigma_{C|MV}$, 
where $\xi_{CM}$, $\xi_{CV}$ and $\sigma_{C|MV}$ are defined analogously 
to $\xi_{MV}$, $\xi_{RV}$ and $\sigma_{V|RM}$ above.  Inserting the 
values from Table~1 of Paper~IV shows that $\xi_{CM}\approx 0$, 
and $\sigma_{C|MV}\approx \sigma_{C|V}=\sigma_C\sqrt{1-\rho_{VM}^2}$:  
the mean color is determined by the velocity dispersion and not by 
the absolute magnitude.  

Passive evolution of the luminosities and colors is incorporated by adding 
the required $z$ dependent shift to $M$ and $C$ after the sizes and velocity 
dispersions have been generated.  

This complete catalog can be used to simulate a magnitude limited 
catalog if we assign each mock galaxy a redshift, assuming a world 
model and homogeneity.  
Let $m_{\rm min}$ and $m_{\rm max}$ denote the apparent magnitude 
limits of the observed sample.  Let $M_{\rm Bright}$ denote the absolute 
magnitude of the most luminous galaxy we expect to see in our catalog.  
Because the luminosity function cuts off exponentially at the bright end, 
we can estimate this by setting $M_{\rm Bright}\approx M_* + 5\sigma_M$.  
This means that the most distant object which can conceivably make 
it into the magnitude limited catalog lies at a luminosity distance of 
about  $d_{\rm Lmax} = 10^{(m_{\rm max}-M_{\rm Bright}-25)/5}$, 
from which the maximum redshift $z_{\rm max}$ can be determined.  
If the comoving number density of mock galaxies is to be independent 
of redshift, we must assign redshifts as follows.  
Draw a random variate $u_1$ distributed uniformly between zero and one, 
and set $d_{\rm Com} = u_1^{1/3}\,d_{\rm Lmax}/(1+z_{\rm max})$.  
The redshift $z$ can be obtained by inverting the 
$d_{\rm Com}(z;\Omega,\Lambda)$ relation.  The apparent magnitude of 
this mock galaxy is 
 $m = M + 5 {\rm Log_{10}}d_{\rm L} + 25 + K(z)$, 
where $K(z)$ is the K-correction.  
If $m_{\rm min}\le m\le m_{\rm max}$, then this galaxy would have been 
observed; add it to the subset of galaxies from the complete catalog 
which would have been observed in the magnitude limited catalog.  

If our simulated catalogs are accurate, then plots of magnitude, 
size, surface-brightness and velocity dispersion versus redshift 
made using our magnitude-limited subset should look very similar 
to the SDSS dataset shown in Figure~12 of Paper~I.  In addition, 
${\rm d}N/{\rm d}z$ in the simulated magnitude limited subset should 
be similar to that in Figure~\ref{nz4}.  Furthermore, any correlations 
between observables in the magnitude limited subset should be just 
like those in the actual SDSS dataset.  If they are, then one has 
good reason to assume that similar correlations measured in the 
complete, rather than the magnitude-limited simulation, represent the 
true correlations between the parameters of SDSS galaxies, corrected 
for selection effects.  In this way, the simulations allow one to 
estimate the impact that the magnitude-limited selection has when 
estimating correlations between early-type galaxy observables.
  
We have verified that our simulated magnitude limited catalogs have 
similar $dN/dz$ distributions to those observed, and the simulated 
$\sigma$ and $R_o$ versus $z$ plots show the same selection cuts at low 
velocities and sizes as do the observed data.  The distribution of 
apparent magnitudes, angular sizes, and velocity dispersions in the 
magnitude limited simulations are very similar to those in the real 
data.  The simulated parameters also show the same correlations at 
fixed luminosity as the data.  Maximum likelihood analysis on the 
simulations produces an estimate of the covariance matrix 
which is similar to that of the data.  Therefore, we are confident that 
our simulated complete catalogs have correlations between luminosity, 
size, and velocity dispersion which are similar to the data.  
%(Because they do not allow for differential evolution of the 
%luminosities, they do not show the redshift dependent $I_o-L$ or 
%$R_o-L$ slopes discussed in Section~\ref{difflz}.)

\section{Composite volume-limited catalogs}\label{compcat}
Our parent sample is magnitude limited; unless accounted for, this 
will introduce a bias into a number of correlations we study in 
this series of papers.  For this reason, we often present results 
measured in a few volume limited subsamples.  
Because of the cuts at both the faint and the bright ends of the 
catalog, each volume-limited subsample used in the main text spans 
only a small range in luminosity.  However, because the galaxies in 
our sample luminosity show little or no evolution relative to the 
values at the median redshift of the sample, we can extend this range 
in either of three ways.  

One method is to construct a composite volume-limited catalog 
by stacking together smaller volume-limited subsamples which are 
adjacent in redshift and in luminosity, but which do not overlap at 
all.  Let $V_i$ denote the volume of the $i$th subsample, and let 
$N_i$ denote the number of galaxies in it.  A conservative approach 
is to randomly choose the galaxies in $V_i$ with probability proportional 
to min$(V_i)/V_i$, where min$(V_i)$ denotes the volume of the smallest 
of the subsamples.  This has the disadvantage of removing much of the 
data, but, because our data set is so large, we can afford this luxury.  
A more cavalier approach is to choose all the galaxies in the largest 
$V_i$, all the galaxies in the other $V_j$, and to generate a set of 
additional galaxies by randomly choosing one of the $N_j$ galaxies 
in $V_j$, adding to each of its observed parameters a Gaussian random 
variate with dispersion given by the quoted observational error, and 
repeating this $N_j\times [{\rm max}(V_j)/V_j -1]$ times.  
A final possibility is to weight all the galaxies in $V_i$ (even 
those which were not in the volume limited subsample) by the inverse 
of the volume in which they could have been observed 
$(V_{\rm max}-V_{\rm min})$.  
We chose the first, most conservative option.  

By piecing together three volume limited subsamples, we were able to 
construct composite catalogs of about $10^3$ objects each.  Because 
the completeness limits are different in the different bands, the 
composite catalogs are different for each band.  In addition, because 
any one composite catalog is got by subsampling the set of eligible 
galaxies, by subsampling many times, we can generate many realizations 
of a composite catalog.  This allows us to estimate the effects of 
sample variance on the various correlations we measure.

\end{document}